\documentclass[review]{elsarticle}
\usepackage{amsmath}
\usepackage{multicol}
\usepackage[font=small,labelfont=bf]{caption}
\usepackage{subcaption}
\usepackage{tablefootnote}
\usepackage{url}

\usepackage{algorithm}
\usepackage{algpseudocode}
\usepackage{hyperref} 
\hypersetup{pdfauthor=Babar Shahzaad}

\usepackage{xcolor}
\begin{document}

\begin{frontmatter}

\title{Resilient Composition of Drone Services for Delivery}

\author[1]{Babar Shahzaad\corref{cor1}}
\ead{babar.shahzaad@sydney.edu.au}
\author[1]{Athman Bouguettaya}
\ead{athman.bouguettaya@sydney.edu.au}
\author[2]{Sajib Mistry}
\ead{sajib.mistry@curtin.edu.au}
\author[3]{Azadeh Ghari Neiat}
\ead{azadeh.gharineiat@deakin.edu.au}
\cortext[cor1]{Corresponding author}
\address[1]{The University of Sydney, Sydney NSW 2000, Australia}
\address[2]{Curtin University, Perth WA 6102, Australia}
 \address[3]{Deakin University, Geelong VIC 3220, Australia}






\begin{abstract}
We propose a novel resilient drone service composition framework for delivery in dynamic weather conditions. We use a skyline approach to select an optimal set of candidate drone services at the source node in a skyway network. Drone services are initially composed using a novel constraint-aware deterministic lookahead algorithm using the multi-armed bandit tree exploration. We propose a heuristic-based resilient service composition approach that adapts to runtime changes and periodically updates the composition to meet delivery expectations. Experimental results prove the efficiency of the proposed approach.
\end{abstract}

\begin{keyword}
DaaS \sep Service selection \sep Service composition \sep Adaptive lookahead \sep Service recomposition \sep Resilient composition
\end{keyword}

\end{frontmatter}


\section{Introduction}

Drones have gained great attention for civil applications from both academic and industrial domains \cite{LIU2019163}. The wide range of applications and services offered by drones show the extensive utilization of drones in various sectors including search and rescue, real-time monitoring, aerial surveillance, structural inspection, and delivery of goods \cite{DBLP:journals/corr/abs-1805-00881} \cite{2}. Several large corporations such as Amazon, DHL, and Google have shown a growing interest in using drones for package delivery \cite{doi:10.1111/drev.10313}. The attractive features of commercial drone delivery are \textit{higher efficiency, cost-effectiveness,} and \textit{higher flexibility} compared to terrestrial transportation \cite{5}.

The \textit{service paradigm} \cite{Bgt2017} provides powerful mechanisms to abstract the \textit{functional} and \textit{non-functional} or \textit{Quality of Service} (QoS) properties of a drone as \textit{Drone-as-a-Service} (DaaS) \cite{8818436}. The functional property of a DaaS describes the delivery of a package from a given source to a destination following a skyway network. The non-functional properties of a DaaS are battery capacity, flight range, payload, and speed. Drone delivery services usually operate in a skyway network to avoid no-fly zones and restricted areas. A skyway network is composed of skyway segments between any two particular nodes following the drone flying regulations such as visual line-of-sight \cite{doi:10.1068/b34066}. The nodes are assumed to be the delivery targets or recharging stations.

The practicality of drone delivery services is limited by a diverse range of \textit{intrinsic} and \textit{extrinsic} factors \cite{8}. The intrinsic factors are the inherited drone's limitations such as limited battery capacity, limited flight range, and constrained payload. The extrinsic factors are related to the drone service environment such as highly dynamic operating environment and constraints on recharging pads at the stations. The maximum flight range of a delivery drone with full payload weight varies from 3 to 33 km \cite{12}. The \textit{battery capacity, speed, payload weight,} and \textit{weather conditions} influence the flight range of a drone \cite{7513397}. 

To the best of our knowledge, existing research mainly focuses on the \textit{scheduling} and \textit{routing} of drones by formulating the problem as Travelling Salesman Problem (TSP) \cite{8488559} and Vehicle Routing Problem (VRP) \cite{7513397}. A single drone routing problem with fuel constraints is studied to minimize the total fuel consumption in \cite{6619438}. The proposed approach is limited to generating routes for only a single drone with a finite number of stations. Detailed analysis on maximizing the profitability and minimizing the drone delivery time is presented in \cite{8}. This approach mainly focuses on battery management of a drone delivery service. However, existing approaches do not consider \textit{recharging constraints} and the \textit{stochastic} nature of drone delivery services. A drone may need multiple times of recharging its battery at intermediate stations for persistent delivery services in long-distance areas. The arrival of the drone services at a recharging station is usually stochastic in nature \cite{venkatachalam2017two}. Each station has usually a finite number of recharging pads. Therefore, the availability of recharging pads may not be guaranteed.  

\textit{Our previous work \cite{10.1007/978-3-030-33702-5_28} is the first to focus on the recharging constraints of drone services using the service paradigm}. In our previous work, we proposed a \textit{novel DaaS composition framework considering the recharging constraints of drone services}. In this context, recharging at intermediate stations leads to the \textit{composition} of DaaS services. The composition provides a means to aggregate the skyway segment services from source to destination \cite{10.1007/978-3-662-45391-9_26}. We formulated the problem of \textit{constraint-aware DaaS composition as a multi-armed bandit tree exploration problem}. We assumed that both the intrinsic and extrinsic factors are \textit{deterministic}, i.e., we know a priori about the available drone services, their QoS properties, and the service environment. Multiple DaaS services instantiated by different drones, operating in the same skyway network at the same time, may cause congestion in the network. We defined \textit{congestion} as the total waiting time require a drone for the availability of recharging pad at a certain station \cite{12}. To avoid congestion within the network, we proposed a lookahead heuristic-based multi-armed bandit approach to compose drone services minimizing the delivery time and cost.

However, \textit{our previous work does not consider the failures in drone services in dynamic weather conditions.} In real-world settings, the drone service environment is \textit{highly dynamic} in nature \cite{BARAKAT2018215}. The QoS properties of drone services may fluctuate due to the changes in the airflow pattern \cite{Mabrouk:2009:QSC:1813355.1813365}. For example, a drone service may arrive late due to strong headwind or may not find a recharging pad available on a certain recharging station due to recharging constraints and stochastic arrival of other drone services. As a result, the drone service may no longer provide the required QoS and fail. Therefore, the initial deterministic composition plan may become non-optimal and need to be \textit{replanned} to deal with changing weather conditions and recharging constraints. \textit{Failure} is a natural phenomenon in service composition. \textit{To the best of our knowledge, no prior work has addressed the failure of drone service composition during the delivery operation.} In this paper, we extend our previous DaaS composition framework \cite{8818436} \cite{10.1007/978-3-030-33702-5_28} by adapting the failures in DaaS composition. Our objective is to propose a \textit{resilient DaaS composition framework}.

We compose the DaaS services and build an initial composition plan using our deterministic approach. The drone services are required to reach certain intermediate stations at a specific time during the delivery operation. The position and time of a drone service are of paramount importance for the smooth execution of the delivery operation. \textit{Failure in DaaS composition means to fail in executing the initial deterministic composition plan.} For example, a drone service $DaaS_1$ needs to reach a recharging station $S_1$ at 02:30 pm. The smooth execution of subsequent drone services depends on the current drone service and the movement of other drone services. The early or late arrival of $DaaS_1$ may affect the other drone services and require to change the initial plan. The early arrival of a drone service at an intermediate recharging station does not necessarily mean to support the initial composition plan. This early arrival may result in long waiting time for the availability of recharging pad. Failure to meet constraints of a composite plan may result in the failure of partial or complete composite drone service.

We propose a \textit{resilient composition of drone services for delivery considering the recharging constraints and uncertain weather conditions}. In this context, resilient means that DaaS composition eventually delivers the package to the destination by adapting failures in the initial deterministic composition plan. The recharging time, weather conditions, and arrival (or departure) of one drone influence the execution plan of other drones at each station. We assume that the available drone services are initially deterministic, i.e., there is a knowledge about the availability of drone services and their QoS values a priori. The real-time delivery operation transforms the deterministic drone services to dynamic and stochastic drone services. The service environment is dynamic and the availability of recharging pads may not be guaranteed. We analyze the local impact of a failed drone service. We then locally recompose the initial composition plan using a novel adaptive lookahead heuristic-based approach. Our proposed approach finds the \textit{best composition plan} from the current position to the next intermediate station where no change to the initial plan has occurred. This process continues until the delivery of the package to the destination.

The main contributions of this paper are as follows:
\begin{itemize}
\item A formal model to represent constraint-aware DaaS services.
\item A Skyline approach for DaaS selection in delivery.
\item A resilient drone service composition approach considering recharging constraints and uncertain weather conditions.
\item A new heuristic-based local service recomposition algorithm using adaptive lookahead approach.
\item A custom drone simulation model for simulating the experiments.
\item An evaluation using a real-world dataset to show the efficiency and effectiveness of the proposed model.
\end{itemize}

\subsection*{Motivating Scenario}

We use a typical drone delivery scenario as our motivating scenario. Drones deliver the packages within Sydney, Australia. Suppose a drone delivery service provider company is planning to deliver a package from \textit{Richmond} to \textit{Cronulla} (89 km). The maximum service distance of a typical delivery drone ranges from 3 to 33 km. The payload weight and wind speed also affect the flight range of a drone. The Bureau of Meteorology (BoM)\footnote{http://www.bom.gov.au/nsw/observations/sydney.shtml} provides the real-time information of wind speed and direction for the Sydney area which helps in determining the flight range of a drone. Multiple times of recharge may be required to serve the delivery request. Avoiding the strong wind areas and the congestion of drones at recharging stations is of paramount importance for time-optimal and cost-effective delivery services.

We construct a skyway network following the Civil Aviation Safety Authority (CASA)\footnote{https://www.casa.gov.au/drones/rules} drone flying regulations such as avoiding no-fly zones and restricted areas. The nodes of the skyway network are the rooftops of high-rise buildings within the Sydney area. Each node can be a \textit{recharging station} or a \textit{delivery target}. Each rooftop has a finite number of recharging pads where a drone can land and recharge. To avoid compatibility issues and present a realistic scenario, we assume that there is no handover of packages at the intermediate stations, i.e., the same drone delivers the package from source to destination. The stochastic arrival of drone services may cause dynamic congestion at certain nodes, i.e., all recharging pads are occupied. Avoiding the congested nodes would result in faster delivery services.

\begin{figure}

    \centering
    \includegraphics[width=\textwidth, height=6cm]{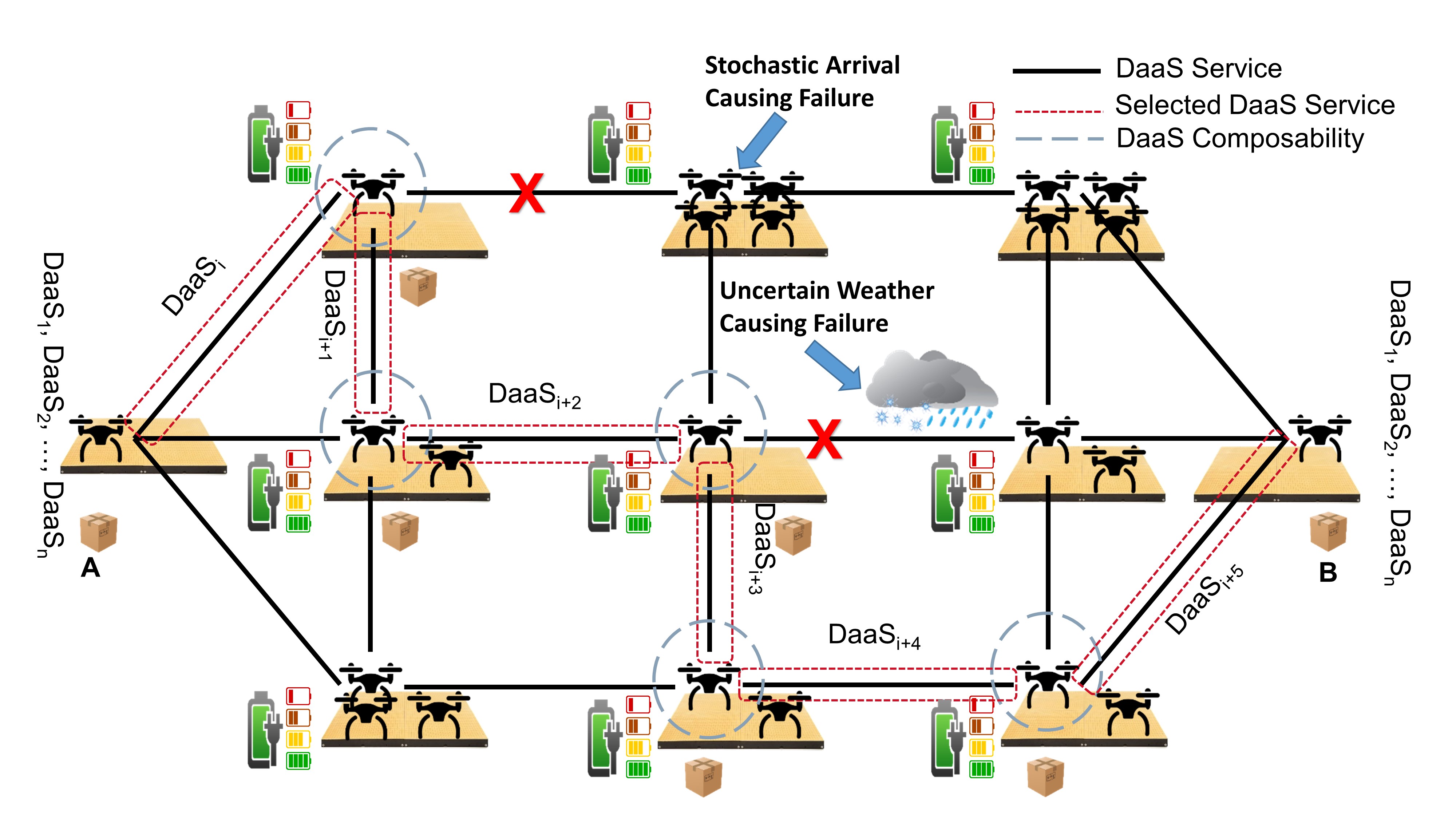}

    \caption{Skyway network for drone-based package delivery considering failures}
    \vspace{-10pt}
    \label{fig1}

\end{figure}

Suppose Yasir needs a package to be delivered within the \textbf{shortest period of time}. If we ignore the uncertainties, e.g., wind effect and the congestion at the stations (i.e., busy recharging pads) and assume that the services are deterministic, the problem would be reduced to finding the shortest path (the composition of skyway segments) from the source to the destination within the skyway network. However, this greedy approach has a higher probability to fail under uncertain conditions in the real-world environment. For example, the initial delivery plan may be highly affected by strong wind in an area or other arriving drone services at a certain station. Fig. \ref{fig1} presents the skyway network for drone delivery with recharging stations, uncertain wind conditions, and failures at intermediate stations.

Our objective is to design a \textit{smart resilient approach} to deal with the effects of failures in the initial deterministic composition plan. This smart resilient approach adapts to the failures automatically, handles the effects of failures, and ensures the on-time package delivery. The brute-force approach considers all the possible compositions to find the best composition plan. However, this approach is highly time-consuming as finding all the possible compositions may produce exponential search space. As a result, we consider the \textbf{local recomposition} approach which updates the initial composition plan when a failure occurs in the initial plan at an intermediate recharging station. We use an \textit{adaptive lookahead heuristic-based approach} which performs the \textit{local optimizations} instead of the \textit{replanning from scratch or global optimization}, i.e., finding the impact of failure in next couple of nodes (relative to the direction of the destination).

\section{Related Work}

To the best of our knowledge, there exists no similar resilient drone service composition approach in the literature considering the dynamic weather conditions. The proposed framework combines concepts from two separate areas: (1) routing and scheduling of drones and (2) failure detection and recovery in composite services. In this section, we overview related work in these two areas.

\subsection{Routing and Scheduling of Drones}

Several studies address the routing and scheduling problems for drone delivery services. Most of the existing research work focuses on using drones in combination with ground vehicles for last-mile delivery. A hybrid framework for ground vehicle and drone was first studied in \cite{MURRAY201586}. They proposed two new approaches for drone-assisted parcel delivery problem to minimize the total delivery time. In the first approach, a drone is launched from the ground vehicle to serve a customer while a ground vehicle is serving another customer. After serving the customer, the drone meets with the ground vehicle in a rendezvous location. In the second approach, the ground vehicle and the drone are separately operated, i.e., the ground vehicle and drone perform dedicated deliveries. \textit{It is concluded that the speed of a drone is an important consideration in determining its flight range} \cite{MURRAY201586}. The proposed approach is tested for small-sized customer instances up to 20. \textit{The proposed hybrid approach requires road access for ground vehicles to make deliveries, i.e., not suitable for remote areas where there is no road infrastructure}.

A single drone routing problem is examined considering multiple refuelling depots in \cite{6619438} where a drone can refuel at any depot. The objective of this study is to minimize the total fuel consumption for visiting all the customers. \textit{It is assumed that drone will never run out of fuel during the journey to a customer}. The problem is modelled using Mixed Integer Linear Programming (MILP) formulation. An approximation algorithm is proposed for solving the problem. The proposed approach is tested for 6 depots and 25 targets only. \textit{The proposed model does not consider the temporal logic constraints}. The proposed approach is restricted to generating delivery routes for only a single drone, i.e., not scalable to be used for multiple drones.

Two multi-trip VRPs problem is proposed considering solely drones to perform deliveries \cite{7513397}. The objective is to minimize delivery time and operational cost. They proposed an energy consumption model based on the relationship between battery capacity and payload weight. Simulated annealing (SA) metaheuristic and MILP solver are used to find sub-optimal solutions for the drone delivery problem. The service area for drone deliveries is limited because all drones are restricted to dispatch from and return to a single depot. The actual flight time, drone speed, and uncertain weather conditions are not taken into account in the proposed model. \textit{The proposed approach does not consider the field recharging which limits the coverage and applicability.}

An energy consumption model is presented for automated drone delivery services in \cite{choi2017optimization}. They assumed that drones can perform multi-package deliveries in a predefined service area. The drone fleet size is optimized by analyzing the impact of payload weight and flight range considering battery capacity. They explore the relationship between four variables (working period, drone speed, demand density of service area, and battery capacity) to minimize the total costs of the drone delivery system. The study indicated that the long hours of operation would benefit both service providers and customers. They found that drone deliveries are more cost-effective in areas with high demand densities. \textit{This study does not consider the dynamic congestion conditions at recharging stations and uncertain weather conditions}.

A scheduling model is presented to support persistent drone delivery services in \cite{SONG2018418}. The relationship between the intrinsic factors such as payload and flight range is considered for the effective use of drone delivery services. Multiple service stations are assumed to replenish batteries of drones during the delivery operation. A MILP formulation is presented to model the problem and solved using a heuristic approach. An exact solution through MILP and a heuristic algorithm are provided. It is assumed that the recharging time at the service station is constant, which is not realistic in practical applications. The flight time is assumed as a function of payload weight. In real-world problems, the flight time depends upon the payload weight, drone speed, and environmental weather conditions such as wind speed and temperature. \textit{The proposed solutions do not take into account the extrinsic factors such as dynamic operating environment, recharging constraints at each station, the influence of one drone’s recharging on other drones, congestion conditions at each station, uncertain weather conditions, and failures in drone delivery services.} Hence, a heuristic-based approach is required which incorporates the aforementioned real-world aspects of drone delivery services.

\subsection{Failure Detection and Recovery in Composite Services}

Many research works discuss the problem of failure detection and recovery in composite services \cite{4580652,1530790,10.1007/978-3-642-17358-5_11,5175901,Saboohi:2012:FRW:2428736.2428787,1452105}. In \cite{4580652}, a service failure recovery approach is presented using subgraph replacement of services containing a failed service. They first represent the composite services as directed graphs. They pre-calculate the subgraphs and then rank them to speed up the recovery process at the time of failure. The subgraph calculation is time-expensive as it considers all possible compositions of all the component services. The subgraph of a failed service is replaced by the best-ranked alternative subgraph. The replacement patterns simply consider the functional and non-functional differences between the new subgraph and replaced subgraph containing the failed service. The proposed approach is highly time-consuming and limited to considering only the sequential digraphs.

A region-based service reconfiguration approach is proposed to repair multiple failed services and satisfy the original end-to-end QoS constraints in \cite{5175901}. A reconfiguration region is composed of one or more failed services. When one or more services in a service composition fail at runtime, they try to replace only those failed services. The proposed approach uses Mixed Integer Programming (MIP) to recompose each region until all regions have a satisfactory composition. Generally, MIP methods are very effective when the size of the problem is small. However, these methods suffer from poor scalability due to the exponential time complexity of the applied search algorithms.

Yu and Lin \cite{1452105} proposed two algorithms to solve service failures. The proposed algorithms compose offline backup service paths for each component service. When a component service incurs a failure, the predecessor of the failed service quickly switches to a backup path to skip the failed service. However, the proposed approach does not consider the QoS in the execution of the composite service. Also, the approach presented can only handle a single point of failure. Because of the dynamic nature of services, the availability of the backing up processes may not be guaranteed when failure happens.

A two-phase approach is proposed for the recovery of failed composite services in \cite{Saboohi:2012:FRW:2428736.2428787}. The two proposed phases are the offline phase and the online phase. In the offline phase, the subgraphs of services are calculated and added to a composite service registry. The offline phase pre-calculations can quicken the replacement. The online phase refers to the execution of composite services. Found subgraphs are ranked according to the semantic description of their component services. The online phase comprises forward and backward approaches. Forward recovery approach attempts to reach the original goal of the composite service by retrying or replacing components and continuing the process. If the forward approach fails to accomplish, the backward approach is applied. The proposed recovery approach does not consider the QoS-awareness capabilities and the dynamism of the execution context environment to adapt the most appropriate recovery strategy.

Recomposition is a naive solution to handle the problem of service execution time failures \cite{saboohi2013automatic}. However, it is extremely time-consuming which is undesired. A repair approach based on planning graphs is proposed as an alternative to recomposition in \cite{10.1007/978-3-642-17358-5_11}. Repair is a form of heuristic and guided partial recomposition. Repair is time-efficient compared to recomposition while generates solutions of similar quality. The proposed approach is restricted to the composition of deterministic services with simple composition requirements. The presented technique does not consider the QoS criteria, which simplifies the problem. In \cite{7543850}, the service composition problem is transformed into a non-deterministic planning problem for creating workflows with contingency plans. The beforehand planning for failures saves execution time. However, the generation of all possible alternative composition plans is a time-consuming process.

A constraint-aware failure recovery approach is proposed to explore the reliability of service composition in \cite{LALEH2018387}. Existing approaches do not consider the constraint verification failures in composite services. They predict failures inside a composite service to reduce the number of service rollbacks upon failure recovery. The proposed solution includes a planning-based service composition approach and a constraint-processing method. The planning-based algorithm constructs constraint-aware composite service plans. The constraint-processing method proceeds with constraint verification in constructed composite service. The proposed approach is restricted to only a small number of possible solutions, i.e., inefficient for a very large number of plans.

An adaptive composition approach is proposed to handle the service changes occurring at runtime, for both repair and optimisation purposes \cite{BARAKAT2018215}. The proposed approach adapts to changes as soon as possible in parallel to the execution process. In this way, the interruption time reduces, the chances of a successful recovery increase, and the most optimal solution is produced according to the current state of the environment. The results show that the proposed approach manages to recover from unexpected situations with minimal interruption, even with frequent changes or in the cases where interference with execution is non-preventable.

The service paradigm is leveraged to abstract the line segment as a service (e.g., a bus service) for multi-modal travel purposes in \cite{9,7862268}. A service composition framework is proposed for composing spatio-temporal line segment services. A novel spatio-temporal A*-based algorithm is proposed to compose the services. It is assumed that the services are deterministic, i.e., time and availability are unknown in advance. A failure-proof composition approach for Sensor-Cloud services is presented in \cite{10.1007/978-3-662-45391-9_26} considering the dynamic features such as position and time. The proposed approach is based on D*Lite algorithm to deal with the changes in QoS of Sensor-Cloud services at runtime.

A spatio-temporal service model is proposed for drone services in \cite{8818436}. A drone delivery function over a line segment in a skyway network is abstracted as a service. A spatio-temporal service model is also proposed for drone delivery services. A spatio-temporal service selection and composition algorithm is proposed to compose line segment services considering QoS properties. The battery capacities and recharging constraints are not considered in the proposed model. A constraint-aware deterministic drone service composition approach is proposed in \cite{10.1007/978-3-030-33702-5_28}. The proposed approach considers the recharging constraints at each station. A skyline approach is presented to select an optimal set of drone services. The drone service composition problem is formulated as a multi-armed bandit tree exploration problem.
A lookahead heuristic-based algorithm is developed to compose the selected services. However, the dynamic service environment, the uncertain weather conditions, and the failure in drone services at runtime are not considered in the proposed approach. \textit{To the best of our knowledge, this paper is the first attempt to model the influence of recharging constraints in a drone service environment and resilient composition of drone delivery services}.

\section{Constraint-Aware System Model for Drone Services}

We propose a constraint-aware system model for drone delivery services. The proposed model includes four main parts: (1) Skyway Network, (2) Drone Services, (3) Effects of Wind Speed and Direction in DaaS, and (4) Constraint-Aware Model for Drone Delivery Services.

\subsection{Skyway Network}

In this section, we describe the structure of the skyway network in which drone delivery services operate. Let $D = \{d_1, d_2,\ldots, d_n\}$ be a set of $n$ drones and $T = \{t_1, t_2,\ldots, t_m\}$ be a set of $m$ delivery targets. The skyway network is represented as an undirected graph $G = (V, E)$, where $V$ is a set of vertices (or nodes) each of which represents a target and $E$ is a set of edges each of which represents a skyway segment service joining any two vertices. We assume that each vertex is also a recharging station. Each node is assumed to have a \textit{finite number of recharging pads}. $B$ is a set of battery capacities for all the drones. The travelling cost and battery consumed in travelling from node $i$ to $j$ are represented by $c_{ij}$ and $b_{ij}$ respectively. The battery consumption of the drone has a proportional relationship with payload weight, the distance travelled by the drone, and the wind speed and direction.

\subsection{Drone Services}

We formally defined a model for drone services in our previous work \cite{8818436}. The proposed model includes the formal definitions of DaaS, DaaS composite service, and DaaS composition problem as follows.

\textbf{Definition 1: Drone-as-a-Service DaaS}. A DaaS is defined as a delivery function of a drone which takes a package from a pickup location to a delivery location (i.e., longitude and latitude) having a start time and an end time and meeting a set of QoS attributes (e.g., flight range). A DaaS is a 3-tuple $<DaaS\_id, DaaS_{f}, DaaS_q>$, where
\begin{itemize}
    \item[$\bullet$] $DaaS\_id$ is a unique drone service ID,
    \item[$\bullet$] $DaaS_{f}$ represents the delivery function of a drone over a skyway segment. The location and time of a DaaS are 2-tuples $<loc_s, loc_e>$ and $<t_s, t_e>$, where
    \begin{itemize}
        \item $loc_s$ and $loc_e$ represent the pickup location and the delivery location,
        \item $t_s$ and $t_e$ represent the start time and the end time,
    \end{itemize}
    \item[$\bullet$] $DaaS_q$ is an n-tuple $<q_1, q_2,\ldots, q_n>$, where each $q_i$ represents a quality parameter of a DaaS, e.g., flight range.
\end{itemize}

\textbf{Definition 2: Composite DaaS Service CS}. A CS is required if a single DaaS service is not able to fulfil the delivery request. A CS is an aggregation of atomic drone services which are combined to satisfy a user's request. The QoS attributes for CS are derived from aggregating the corresponding QoS attributes of atomic DaaS services. For example, the delivery cost of a CS is the summation of delivery costs of all atomic DaaS services in a CS. A CS is a 3-tuple $<CSID, CSF, CSQ>$, where
\begin{itemize}
    \item[$\bullet$] $CSID$ is a concatenation of each component DaaS $DaaS_i \in CS$, i.e.,  $CSID = concat(DaaS_i.id)$
    \item[$\bullet$] $CSF$ is a set of functions $\{f_1(DaaS_1), f_2(DaaS_2), \ldots, f_n(DaaS_n)\}$, where each $f_i$ represents the function of corresponding component DaaS $DaaS_i \in  CS$
    \item[$\bullet$] $CSQ$ is an m-tuple $<Q_1, Q_2, \ldots, Q_m>$, where each $Q_j$ denotes an aggregated value of $j^{th}$ quality parameter of component DaaS $DaaS_i \in CS$.
\end{itemize}

\textbf{Definition 3: DaaS Composition Problem}. For a given set of DaaS $S_{DaaS} = \{DaaS_1, DaaS_2,..., DaaS_n\}$ services in a skyway network, the DaaS composition problem is to compose the services for delivering a package from a pickup location to a delivery location in minimum time.

\subsection{Effects of Wind Speed and Direction in DaaS}

The wind is a major environmental factor affecting the drone's performance and flight behaviour \cite{10.1007/978-3-319-99996-8_16}.
The wind effect that causes the drone to drift in a certain direction is studied in \cite{selecky2013wind}. They designed a method based on a modified accelerated A* algorithm to take the wind effects into account and generate reachable states. It is assumed that the wind is constant which does not reflect the real-world scenarios. A deadline-constrained routing scheme is presented for delivery drones in \cite{BNCSS102}. The objective of this study is to minimize the energy consumption under wind conditions.

We consider the effects of wind speed and direction in dynamic weather conditions. Highly random nature of wind speed and direction (i.e., headwind and tailwind) greatly influences the battery consumption rate and flight range of the drone \cite{doi:10.2514/6.2005-6951} \cite{VAZQUEZ2017304}. We present a model to determine the impact of wind speed and direction on the travel time of a drone. The travel time of a drone increases with headwind and reduces with the tailwind. We calculate the effects of wind speed and direction on travel time using a method in \cite{10.2307/43943662} for a drone travelling from node $i$ to $j$ as follows.
\begin{align}
\begin{split}\label{eq:1}
    \delta &= \theta_{ij} - \theta_{WS}
\end{split}\\
\begin{split}\label{eq:2}
    A &= WS.\cos(180-\delta)
\end{split}\\
\begin{split}\label{eq:3}
    C &= WS.\sin(180-\delta)
\end{split}\\
\begin{split}\label{eq:4}
    B &= \sqrt{AS^2-C^2}
\end{split}\\
\begin{split}\label{eq:5}
    GS &= A + B \\
    &= WS.\cos(180-\delta) + \sqrt{AS^2-WS^2.\sin^2(180-\delta)}
\end{split}\\
\begin{split}\label{eq:6}
    T_{ij} &= \frac{d_{ij}}{GS} 
\end{split}
\end{align}
where,
\begin{multicols}{2}
\begin{itemize}
    \item{\makebox[0.5cm]{$\theta_{ij}$\hfill} = bearing from node $i$ to $j$}
    \item{\makebox[0.5cm]{$\theta_{WS}$\hfill}  = wind bearing}
    \item{\makebox[0.5cm]{$\delta$\hfill}  = course correction angle}
    \item{\makebox[0.5cm]{$WS$\hfill}  = wind speed}
    \item{\makebox[0.5cm]{$A$\hfill}  = headwind/tailwind. When $|\delta| < 90$, $A$ is negative and denotes headwind. When $90 < |\delta| \leq 180$, $A$ is positive and denotes tailwind.}
    \item{\makebox[0.5cm]{$C$\hfill}  = wind adjustment angle}
    \item{\makebox[0.5cm]{$B$\hfill}  = wind adjustment angle}
    \item{\makebox[0.5cm]{$AS$\hfill}  = air speed}
    \item{\makebox[0.5cm]{$GS$\hfill}  = ground speed}
    \item{\makebox[0.5cm]{$d_{ij}$\hfill}  = distance between node $i$ and $j$}
    \item{\makebox[0.5cm]{$T_{ij}$\hfill}  = travel time from node $i$ to $j$}
\end{itemize}
\end{multicols}

\subsection{Constraint-Aware Model for Drone Delivery Services}

In this section, we first present our previous constraint-aware DaaS composition model for drone delivery services. The constraint-aware composition means to compose the drone services knowing the availability of recharging pads at intermediate stations and the arrival of other drone services. In our previous work \cite{10.1007/978-3-030-33702-5_28}, we assume all the drone services and service environment are deterministic, i.e., the QoS attributes of drone services, the availability of recharging pads, and the trajectory of other drone services are all known beforehand. Our objective was to compose the drone services avoiding the congested recharging stations and delivering the packages in the shortest time. However, such an assumption of the deterministic service environment is not realistic in practice. The QoS may fluctuate and fail due to the dynamic nature of drone services and changing wind patterns. We relax the assumption of the deterministic service environment. We consider that the service environment is stochastic and flight time may vary with the changing wind conditions and the arrival of other drone services.

We compose the drone services to generate an initial service composition plan using our previous deterministic approach. Different types of drones have varying payloads, flight ranges, and battery capacities. There is a constraint that the same drone delivers the package from source to destination. A drone can either \textit{recharge, wait, or travel from one station to the next station}. The deterministic approach estimates the \textit{arrival time, waiting time,} and \textit{recharging time} of each drone at a specific recharging station. The initial composition plan adapts to the failures \textit{dynamically} occurred at runtime. Here, failure means the late or early arrival of drones than the scheduled arrival in the initial plan. This failure may have a \textit{cascading effect} to the execution of subsequent drone services, thus affecting the initial composition. Therefore, a resilient DaaS composition framework is required to ensure the on-time delivery of drone services.

\section{Drone Service Selection using Skyline Approach}

The first step to compose drone services is the selection of appropriate candidate services. For this purpose, we consider several drone services from multiple service providers. The QoS properties of drone services distinguish among functionally equivalent services. Some of the available drones may not carry the package because of its higher weight. Therefore, we use the difference between the payload capacity of the drone and package weight to filter out the candidate drone services. We use skyline approach \cite{914855}\cite{5552749} to further reduce the number of candidate drone services by selecting only the non-dominated services. Skyline computation \textit{speeds up} the service selection process and selects the services with \textit{best QoS attributes}. Skyline approach is also used to deal with the uncertainty of service in the process of selection \cite{Papadias:2003:OPA:872757.872814}. A multi-attribute optimization technique, called service skyline computation, guarantees to provide the best user-desired service providers \cite{Yu2012}.

For a given set $DaaS = \{DaaS_1, DaaS_2, \ldots, DaaS_n\}$ of functionally similar drone services and a set $Q = \{q_1, q_2, \ldots, q_m\}$ of QoS attributes, we present formal definitions of drone service domination and service skyline as follows.

\textbf{Definition 4: Drone Service Domination}. The domination relationship between a drone service $DaaS_i \in DaaS$ and another drone service $DaaS_j \in DaaS$ is defined as $DaaS_i \prec DaaS_j$, if $\forall q_k \in Q $, $q_k(DaaS_i) \preceq q_k(DaaS_j)$, and $\exists q_l \in Q $, $q_l(DaaS_i) \prec q_l(DaaS_j)$ where $\prec$ denotes better than and $\preceq$ denotes better than or equal to relationship.

\textbf{Definition 5: Service Skyline}. The service skyline comprises a set of drone services, denoted by $SKY_{DS}$, that are not dominated by any other drone service, i.e., $SKY_{DS} = \{DaaS_i \in DaaS | \neg\exists DaaS_j \in DaaS : DaaS_j \prec DaaS_i\}$.

We compute the skyline using the following three QoS properties: (1) \textit{flight time} (in minutes) represents the time duration a drone can fly with battery charged to its capacity, (2) \textit{flight range} (in kilometres) represents the distance a drone can travel with full capacity charge, and (3) \textit{recharging time} (in hours) for 0 to 100\% recharge. We use \textit{Block Nested Loop (BNL)} algorithm \cite{914855} for skyline computation. The non-dominated skyline services are obtained by repetitive scanning of the candidate drone services. The \textit{BNL} algorithm can be used for any dimensionality without requiring any indexing or storage. It performs well most of the time for dealing with our low dimension and small domain range data. Table \ref{tab:table1} presents an example of skyline computation for functionally similar drone services that are differed in QoS properties. For instance, a drone service $DaaS_7$ dominates another drone service $DaaS_9$ according to aforementioned domination relationship. The ``Is skyline?" column illustrates the outcome of skyline computation.

\begin{table}[t]
\centering
\scriptsize
\caption{A set of functionally similar Drone services}

\label{tab:table1}
\resizebox{\columnwidth}{!}{\begin{tabular}{|c|c|c|c|c|c}
\hline
 \textbf{Drone service} &  \textbf{Flight time (min)} & \textbf{Flight range (km)} & \textbf{Recharging time (hours)} & \textbf{Is skyline?} \\

\hline

$DaaS_1$ &  20  & 0.8 & 1.5 & No \\ \hline

$DaaS_2$ & 20 & 56 & 2 & Yes \\ \hline

$DaaS_3$ & 25 & 8 & 1 & Yes \\ \hline

$DaaS_4$ & 30 & 7 & 1.5 & No \\ \hline

$DaaS_5$ & 20 & 1.6 & 1.5 & No \\ \hline

$DaaS_6$ & 18 & 0.8 & 1.5 & Yes \\ \hline

$DaaS_7$ & 120 & 100 & 2 & Yes \\ \hline

$DaaS_8$ & 20 & 3 & 1 & Yes \\ \hline

$DaaS_9$ & 27 & 7 & 1 & No \\ \hline

$DaaS_{10}$ & 40 & 1.9 & 1.5 & No \\ \hline

$DaaS_{11}$ & 22 & 5 & 1.5 & Yes \\ \hline

$DaaS_{12}$ & 24 & 8 & 1.5 & Yes \\ \hline

\end{tabular}}
\end{table}

\section{Resilient Drone Service Composition Framework}

\begin{figure} 

    \centering
    \includegraphics[width=\textwidth]{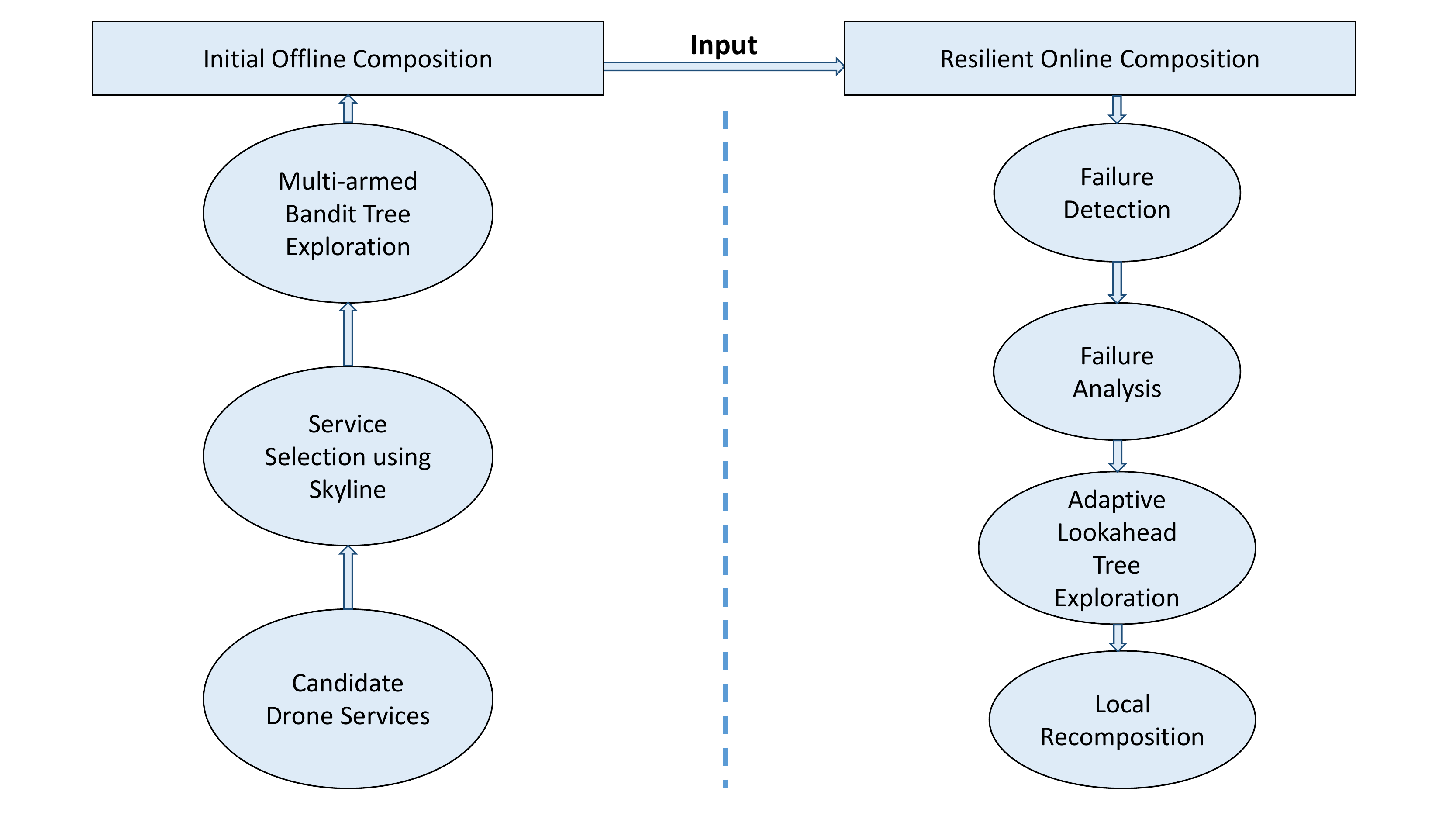}

    \caption{Resilient Drone Service Composition Framework}

    \label{fig6}

\end{figure}

We divide the resilient drone service composition framework into two categories: (1) Constraint-Aware Drone Service Composition using Lookahead and (2) Resilient Drone Service Composition using Adaptive Lookahead. Fig \ref{fig6} presents an overview of the resilient drone service composition framework. The initial offline composition is provided by constraint-aware drone service composition in a deterministic fashion. While the resilient online composition is carried out to handle the dynamic failures in the initial offline composition at runtime.

\subsection{Constraint-Aware Drone Service Composition using Lookahead}

\begin{figure} [t]

    \centering

    \includegraphics[width=\textwidth]{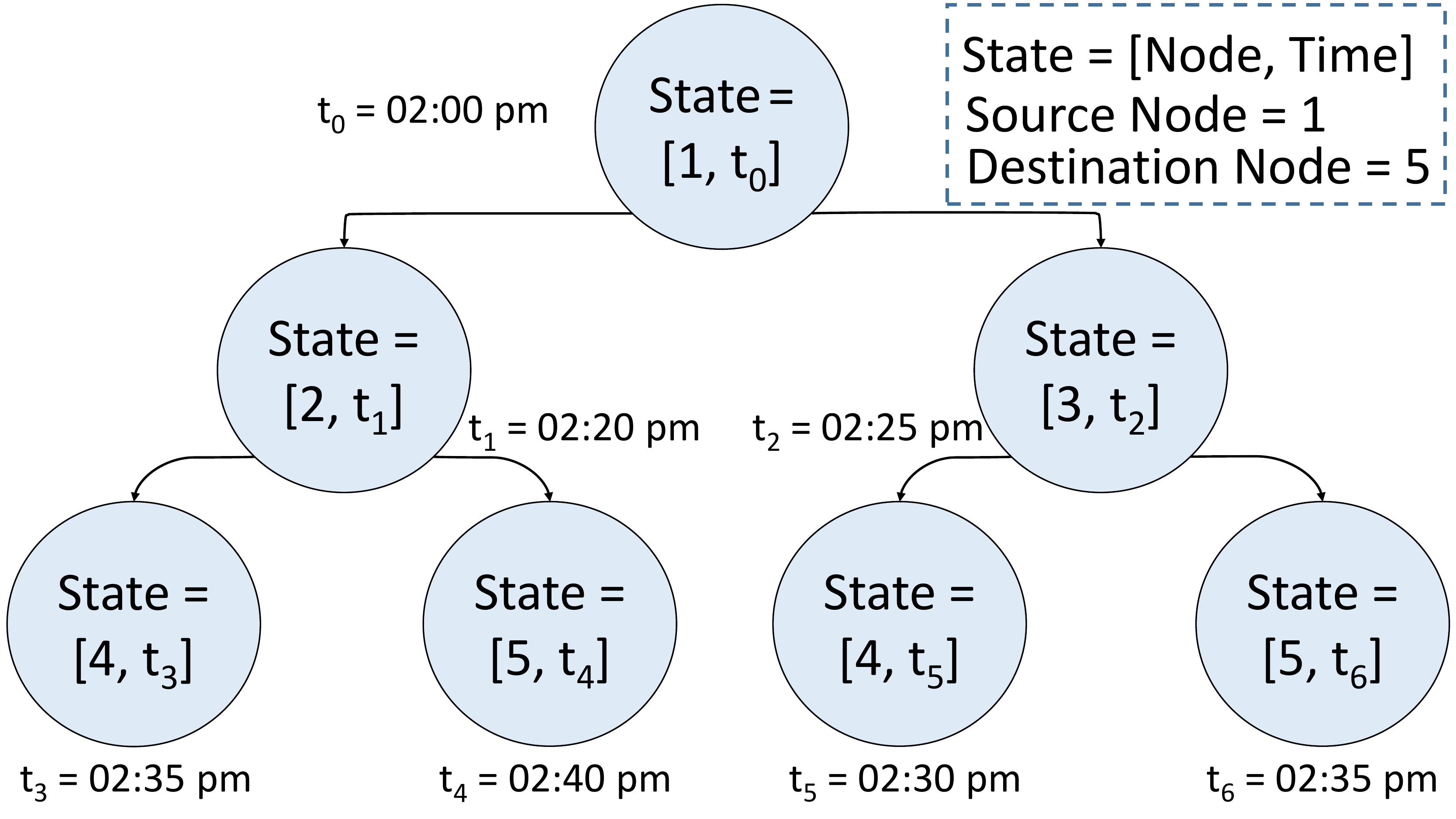}

    \caption{An example of a state tree}
    
    \label{fig2}

\end{figure}

We formulate the constraint-aware drone service composition as the multi-armed bandit tree \cite{Coquelin:2007:BAT:3020488.3020497} exploration problem. In multi-armed bandits, an arm denotes an action or a choice which is initially unknown to the player. If the arms are deterministic, i.e., known beforehand, the problem would be reduced to the selection of arms with the highest reward. We assume that the drone services and the services environment are initially deterministic. Our target is to maximize the reward by selecting optimal arms. A drone can take the following set of actions at each station: recharge, wait, or travel from one station to the next. These actions generate a large set of possible states. Fig. \ref{fig2} presents an example of a temporal state tree. 
We formally define a state as follows:

\textbf{Definition 5: State}. A state is a tuple of $<NodeID, TimeStamp>$, where
\begin{itemize}
    \item[$\bullet$] $NodeID$ is a unique node identifier,
    \item[$\bullet$] $TimeStamp$ represents the arrival time of drone at a certain node.
\end{itemize}

For the sake of simplicity, we consider that the states are known beforehand. In case of immediate state selection, the temporal optimal neighbour state may lead to a non-optimal state, e.g., long waiting time due to congestion at the next station.

The selection and composition of optimal drone services from a large number of candidate services is a challenging task. The uncertainty is the main issue in a DaaS composition. In many cases, an immediate optimal service may lead to a non-optimal service. For example, we have a skyway network where node 1 is the source node and node 5 is the destination node. Here we find a temporal optimal neighbour leading to a non-optimal state. Temporal optimal means taking towards destination faster. As shown in Fig. \ref{fig2}, the service of state [2, $t_1$] is optimal but the overall delivery time is more compared to state [3, $t_2$]. This uncertainty can cause long delays for drones to deliver packages. Looking for all possible service compositions or deep tree exploration is not computationally feasible to find the best composition. The time complexity for such problems is exponential. Hence, we need a heuristic-based solution to find the optimal composition of drone services.

\begin{figure} [t]
   \centering
    \includegraphics[width=\textwidth]{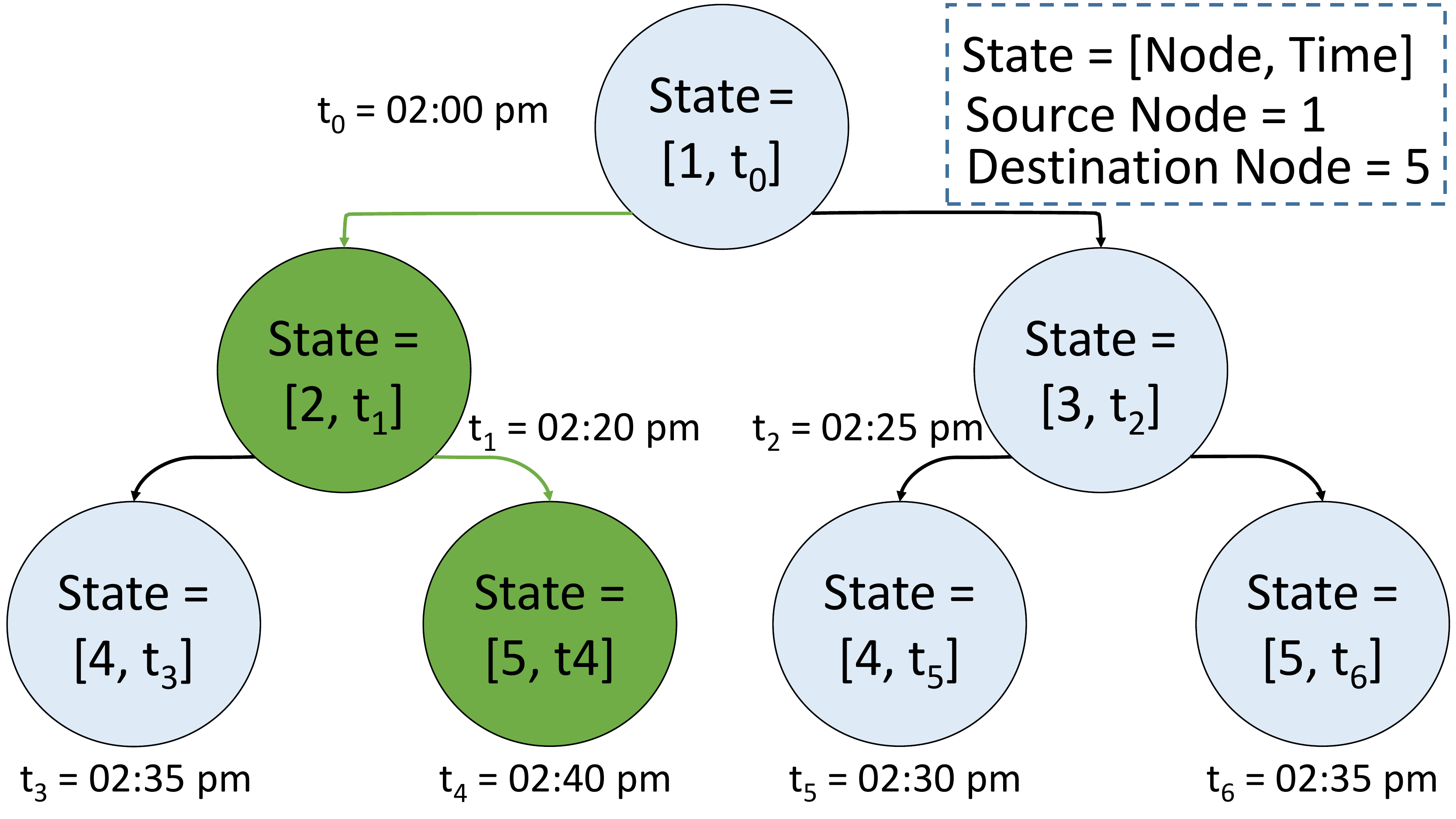} 
    \caption{State selection without lookahead}
    \label{fig3}
\end{figure}

\begin{figure} [t]
    \centering
    \includegraphics[width=\textwidth]{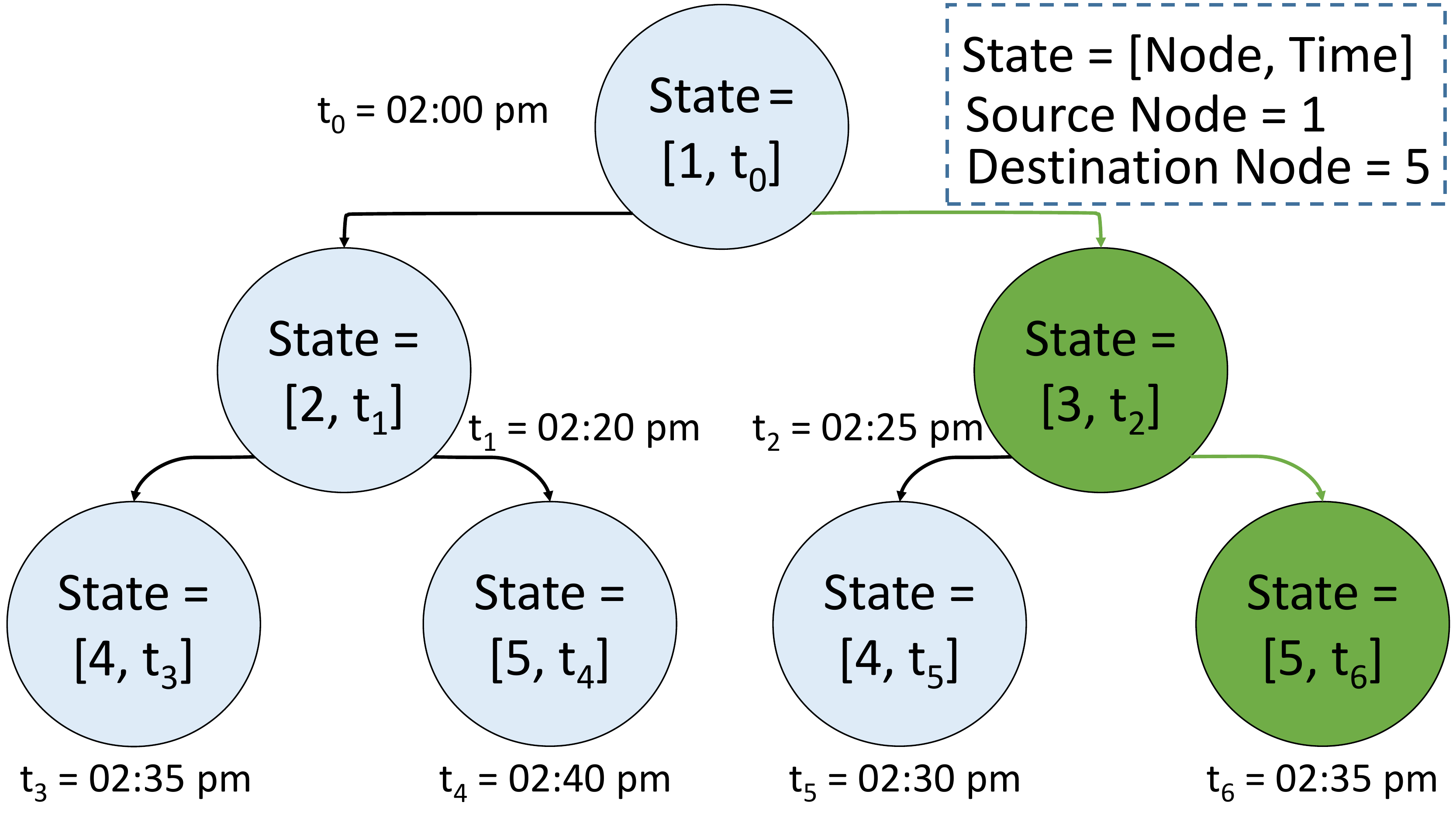} 
    \caption{State selection with one lookahead}
    \label{fig4}
\end{figure}

We propose a lookahead heuristic-based solution to the multi-armed bandit tree exploration problem. The selection of optimal actions in a DaaS composition is performed by looking ahead of neighbour services. We consider the current waiting time, expected waiting time, and flight time to the destination for selection of optimal drone services. The term lookahead means considering the next-to-adjacent states while making the state selection decision. Fig. \ref{fig3} and \ref{fig4} illustrate the difference between without lookahead and with one lookahead based service (state) selection. Without lookahead considers only the neighbour optimal states which leads to an overall non-optimal solution. Using lookahead heuristic provides more information to select the overall optimal states. We build our initial composition plan using the aforementioned lookahead strategy. 
But, this approach does not take into account the runtime failures in the execution of the initial composition plan such as uncertain weather conditions. We need a resilient composition approach for drone services to ensure the in-time package delivery.

\subsection{Resilient Drone Service Composition using Adaptive Lookahead}

The underlying initial DaaS composition approach is formulated as a multi-armed bandit problem \cite{10.1007/978-3-030-33702-5_28}. Multi-armed bandits are a special type of sequential decision problems which demonstrate exploration and exploitation trade-offs and produce maximum rewards under uncertainty \cite{villar2015multi}. The exploration refers to trying each possible action to find an optimal reward. In contrast, exploitation refers to trying the actions that are believed to provide higher payoffs in the future. We focus on the constraints at recharging stations and dynamic weather conditions. However, multi-arm bandits are generally proposed for tree-based search exploration in the context of combinatorial optimization \cite{7969356}. An exact approach such as MILP does not naturally fit to solve such exploratory optimization \cite{RADMANESH2016149}. MILP approaches are usually applied in solving deterministic linear optimization problems \cite{lin2012review}. While heuristic-based lookahead \cite{NIPS2017_6785}, genetic algorithm \cite{Berro2010}, and tabu search \cite{soykan2016hybrid} are usually used for exploratory optimization problems. We focus on the adaptive lookahead heuristic-based approach which is typically used to solve combinatorial multi-armed bandit problems \cite{10.1145/1273496.1273587}. The heuristics are widely used in multi-armed bandit literature and provide substantially more efficient solutions than traditional optimization approaches \cite{NIPS2019_9592}. Therefore, we focus on exploring a heuristic-based solution for the composition of drone services.

The stochastic arrival of other drone services at intermediate stations and the changes in wind pattern influence the initial composition plan. As a result, the established composition plan may become non-optimal and fail. Such failures may impact on the initial composition in two ways: (1) local impact (2) global impact. The term local impact means the effect of failure propagates to a certain number of recharging stations. The rest of the plan is still recoverable. The term global impact refers to the propagation of failure effect till the destination.

\begin{figure} [t]

    \centering
    \includegraphics[width=\textwidth, height=8cm]{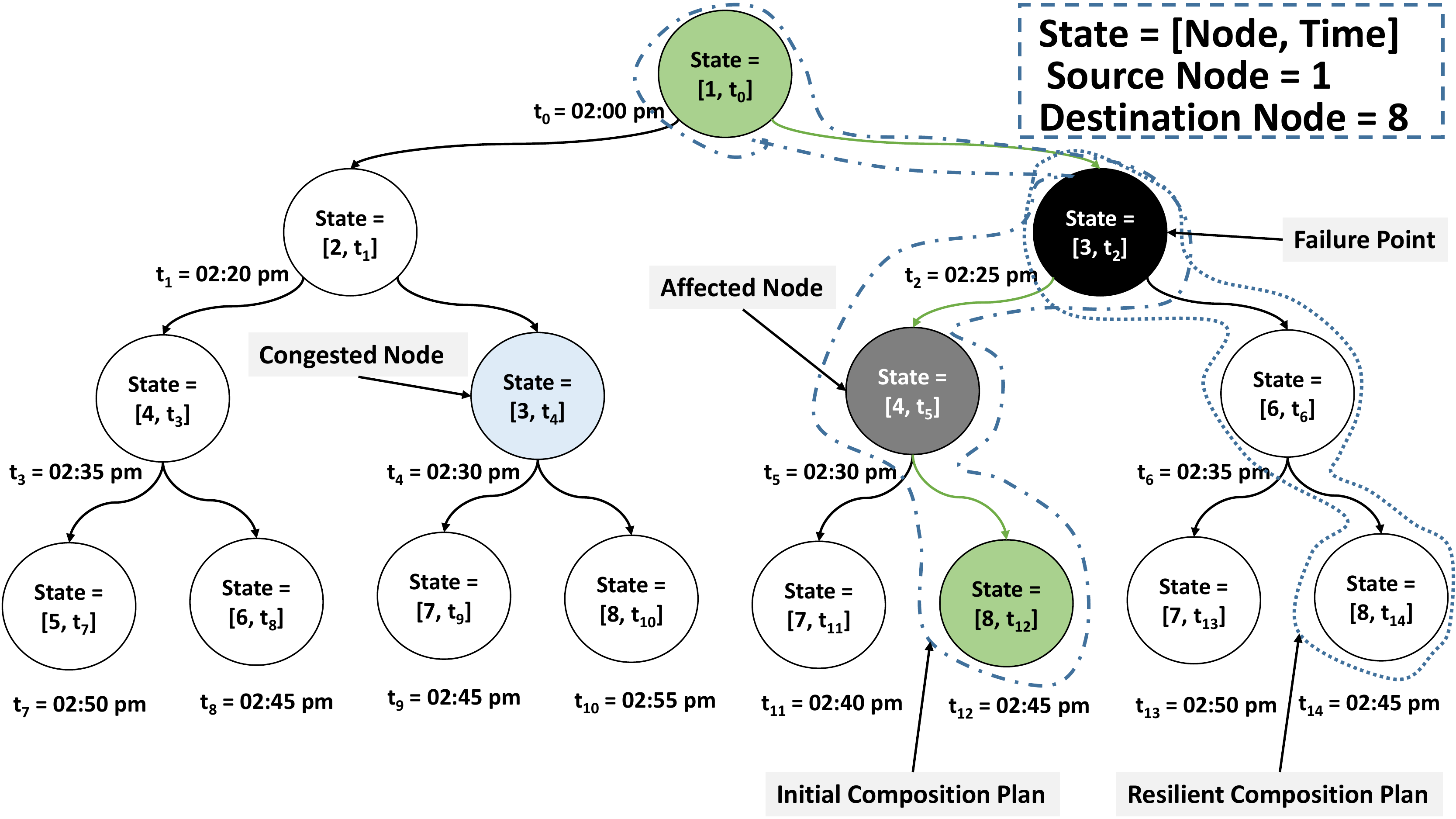}

    \caption{State selection using adaptive lookahead}

    \label{fig13}

\end{figure}

\begin{figure} [t]

    \centering
    \includegraphics[width=\textwidth, height=8cm]{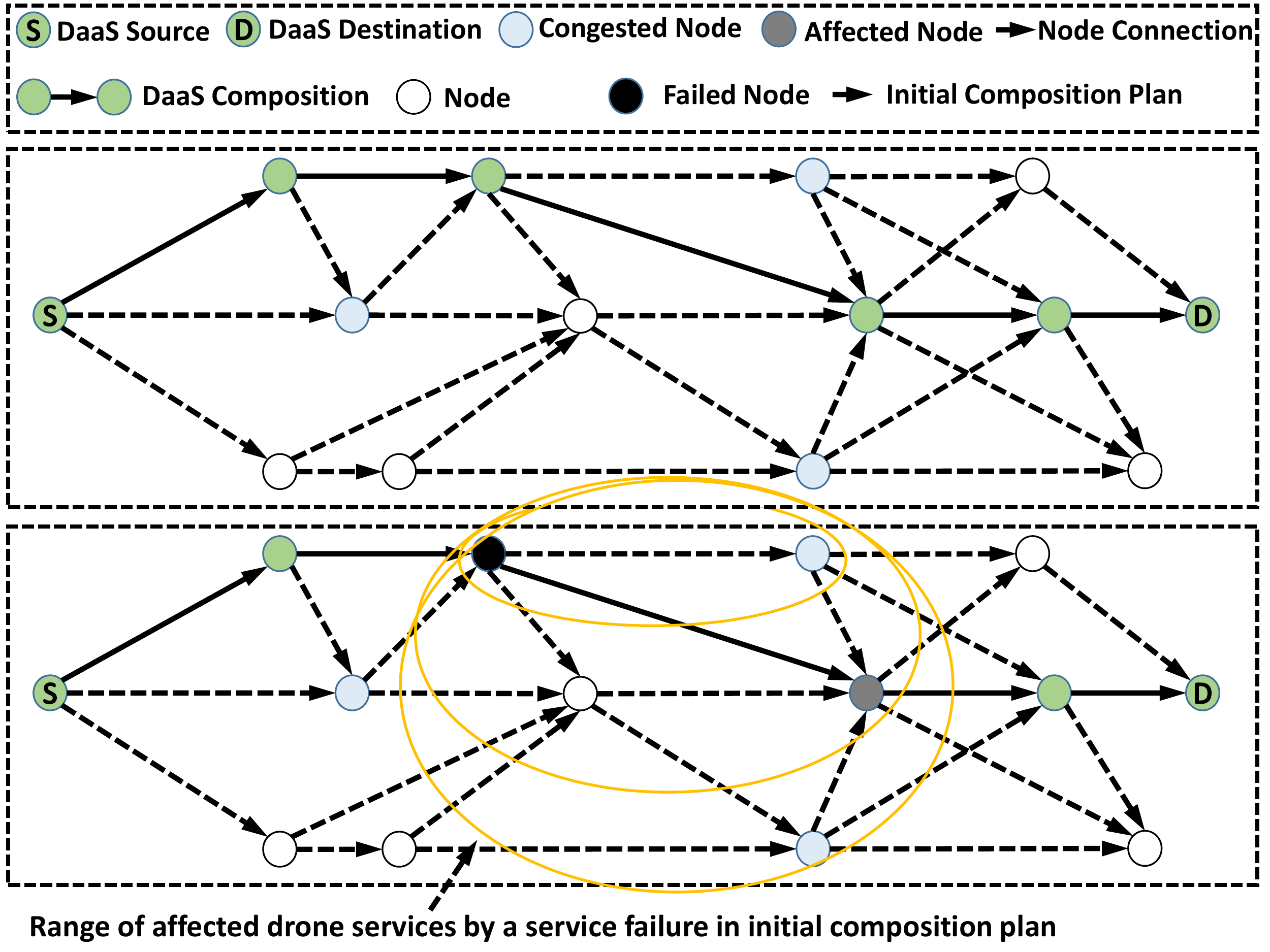}

    \caption{An example of initial composition plan failure and its impact}

    \label{fig5}

\end{figure}

We propose a resilient drone service composition using adaptive lookahead heuristic-based approach. The resilient means that the delivery operation is successfully carried out even the established composition plan adapts to the failures. We require a lookahead algorithm to handle time-varying constraints and weather conditions. The adaptive lookahead performs lookahead according to the type of failure occurred rather than a fixed number of lookaheads. There is only one difference between an adaptive lookahead and a standard lookahead algorithm: the distance in adaptive lookahead is no longer a fixed length but varies with the propagation effect of the failures. Once the adaptive lookahead finds the distance of failures, we locally recompose the drone services from the current station to the next failure-free station. The state selection using adaptive lookahead approach is shown in Fig. \ref{fig13}. The selective states of initial composition plan are represented by green colour. Failure occurs at node 3 which requires the recomposition of services. We recompose the services until the next state where no change is observed. Fig. \ref{fig5} presents an example of initial service composition, failed service, and its impact on other services. We first compute an initial offline composition plan from source to destination, denoted by a sequence of solid line arrows connecting green colour nodes. The initial composition plan avoids the congested recharging stations (yellow colour nodes) for faster delivery services. A failed service is represented by a red colour node and its impact on the next services in the initial plan is shown by orange colour nodes. The formal definitions of failure, service failure, and resilient service composition are given as follows.

\textbf{Definition 6: Failure}. A failure is defined as the deviation from expected (specified) behaviour. In some cases, the failure may result in the termination of the ability to perform the required function.

\textbf{Definition 7: Service Failure}. Service failure is defined as an event that occurs when the delivered service deviates from the correct service. For example, a drone service $DaaS_i$ is specified to reach a station $Station_j$ at $04:00$ pm. If $DaaS_i$ reaches at $Station_j$ before or after $04:00$ pm, we say that the service is failed.

\textbf{Definition 8: Resilient Service Composition}. Resilience refers to the ability or capacity of a system to adapt to dynamic changes (failures) without deviating from the expected behaviour. Resilient service composition is a mechanism for handling the failures occurred at runtime. When one or more services fail at runtime, the resilient service composition approach locally or partially recomposes the failed services to deliver the expected behaviour.

Fig \ref{fig6} illustrates the process of resilient drone service composition. We first execute the initial offline composition plan. The failure detection module periodically checks for any failures at each station. Each drone service has a certain deadline for each station defined in the initial offline plan. We compare the current arrival time of a drone service with an expected arrival time given by the established plan. In the case of the early or late arrival of a drone service, the failure is detected. The failure analysis module finds the number of services affected due to failure. The failure may affect the execution of a couple of next drone services. The adaptive lookahead tree exploration module guarantees the exploration of all possible alternatives to the failed service. Finally, we locally recompose the explored alternatives to mitigate the effect of failure. The recomposition of drone services at the intermediate station obtains an optimal composite service in minimal computational time.

\begin{algorithm}[t]
 \caption{Resilient Drone Service Composition Algorithm}\label{alg:algorithm1}
    \begin{algorithmic}[1]
    \Procedure {Execute\_Init\_Plan} {$InitComp$}
    \State $DaaS_{cur} \gets InitComp [start]$
    \State $DaaS_{dst} \gets InitComp [end]$
    \While{$DaaS_{cur} \ne DaaS_{dst}$}
    \State Execute initial composition plan
    \State Monitor the execution to find the failed services
    \State $fd \gets$ failure\_detection ($DaaS_{cur}.t_e, curTime$)
    \If{$fd$}
    \State $DaaS_{affected} \gets$ failure\_analysis ($InitComp, DaaS_{cur}$)
    \State $LocalComp \gets$ recompose ($InitComp, DaaS_{cur}, DaaS_{affected}, curTime$)
    \State $InitComp \gets$ update\_plan ($InitComp, LocalComp, DaaS_{cur}$)
    \EndIf
    \State $DaaS_{cur} \gets InitComp [next\_DaaS]$
    \EndWhile
    \EndProcedure
    \end{algorithmic}
\end{algorithm}


\begin{algorithm}[t]
 \caption{Failure Analysis}\label{alg:algorithm3}
    \begin{algorithmic}[1]    
    \Procedure{failure\_analysis}{$InitComp, DaaS_{cur}$}
    \State $failedDaaS \gets 1$
    \State Find first congested node $CongNode$ from $DaaS_{cur}$ to the destination
    \If{$CongNode$}
    \State $DaaS_{affected} \gets$ compute number of services from $DaaS_{cur}$ to $CongNode$
    \EndIf
    \For {each $DaaS \in InitComp$ from $DaaS_{cur}$}
    \State $td \gets$ compute time difference between actual and expected $DaaS$
    \If{$td \geq 0$}
    \State $failedDaaS \gets failedDaaS + 1$
    \Else
    \State break
    \EndIf
    \EndFor
    \State \Return min($DaaS_{affected}, failedDaaS$)
    \EndProcedure
    \end{algorithmic}
\end{algorithm}

\begin{algorithm}
 \caption{Recomposition of Drone Services}\label{alg:algorithm4}
    \begin{algorithmic}[1]    
    \Procedure{recompose}{($InitComp, DaaS_{cur}, Ld_{adapt}, curTime$)}
    \State $srcLocal = DaaS_{cur}$
    \State $dstLocal = InitComp[DaaS_{cur}.index + Ld_{adapt}]$
    \State $startTime = curTime$
    \State $newComp = $ find\_optimal\_comp ($G, RP, D, srcLocal, dstLocal, w,$ $ Ld_{adapt}, WS, \theta_{WS},startTime $)
    \State \Return $newComp$
    \EndProcedure
    \end{algorithmic}
\end{algorithm}

Algorithm \ref{alg:algorithm1} provides the details of the proposed approach as follows. The algorithm generates a resilient composition of drone services using an initial composition plan as input. The first and last component services in the initial plan are the source and destination locations (Lines 2-3). We execute the initial plan and monitor periodically for any failure at runtime (Lines 4-6). The initial plan is executed smoothly until a failure is detected (Line 7). The actual arrival time at each station is compared with the expected arrival in the initial plan. If a failure is detected, the failure analysis algorithm computes the affected (i.e., failed) drone services (Line 9). Algorithm \ref{alg:algorithm3} presents the details of the failure analysis algorithm. We find the first congested node in the initial plan from the failed service until the destination. If a congested node is found, we simply compute the number of services from the current failed service to the congested node. Moreover, we find the first unaffected service from the failed service until the destination. We calculate the difference between the first failed service and unaffected service in the initial plan. The minimum of distance (i.e., nodes) is selected from the congested node and unaffected service. We consider the congested node for failure analysis because the delay of service failure results in less waiting time at a congested node. For example, a service failure causes 15 minutes delay to the initial composition plan. Let's assume that there is a congestion node in the initial plan ahead of failed service. The waiting time on the congested node is 25 minutes for the availability of recharging pad. In such a case, the waiting time will be reduced to 10 minutes because of 15 minutes delay from failed service. The adaptive lookahead distance is equivalent to the number of affected drone services for exploration of all possible alternatives. We recompose the services from the failed position to next unaffected drone service using recompose algorithm (Line 10). The details of recompose algorithm are given in Algorithm \ref{alg:algorithm4}. The recompose algorithm composes the services locally by calling the $find\_optimal\_comp$ function which is same as our drone service selection and composition algorithm in \cite{10.1007/978-3-030-33702-5_28}. Finally, the new locally composed drone services update the inconsistent affected services in our initial composition plan. This process continues until the package is delivered to the destination. An alternative to the use of local recomposition is to replicate the delay in the initial service composition till the destination. This alternative approach may result in longer delays, and in some cases, the package may not be delivered.

\section{Experiments and Results}

We evaluate the effectiveness of the proposed resilient drone service composition approach in this section. A set of experiments are conducted to assess the performance of the proposed approach. We compare the proposed approach with a baseline (i.e., Brute-Force) approach and a without lookahead approach. The most important features of the drone delivery services are the shortening of the delivery time and cost reduction. The delivery cost is a function of drone travelling distance. Therefore, we mainly focus on three evaluation metrics: (1) \textit{delivery time}, (2) \textit{computation time}, and (3) \textit{distance travelled}.
All the experiments are conducted on an Intel Core i9-9900X processor (3.50 GHz) with 32.0 GB memory under Windows 10. Python is used to implement the algorithms.

\subsection{Experimental Setup}

Simulation tools offer a faster, cost-effective, and safe approach to assess the performance of possible solutions before physical testing. There exists several simulators for drones, e.g., AirSim \cite{10.1007/978-3-319-67361-5_40}, Gazebo \cite{1389727}, and JMavSim \cite{babushkin2018jmavsim}. These simulators are not specifically designed for drone delivery services over skyway networks with recharging stations. For example, AirSim does not model drone energy consumption in dynamic environments \cite{DBLP:journals/corr/abs-1906-00421}. Energy is a scarce resource in drones that affects the entire delivery operation. The AirSim platform does not implement payload effects on the power consumption of the drone. The failures in delivery services are not considered in AirSim. The skyway network for drone delivery services is also not a part of the AirSim platform. As the centre of our paper is the drone-based delivery platform, we implement a custom drone-based delivery simulation model for the experiments. In future, we plan to deploy a skyway network and delivery management framework on AirSim for greater reachability to the research community.

\begin{figure} [t]

    \centering
    \includegraphics[width=\textwidth, height=6cm]{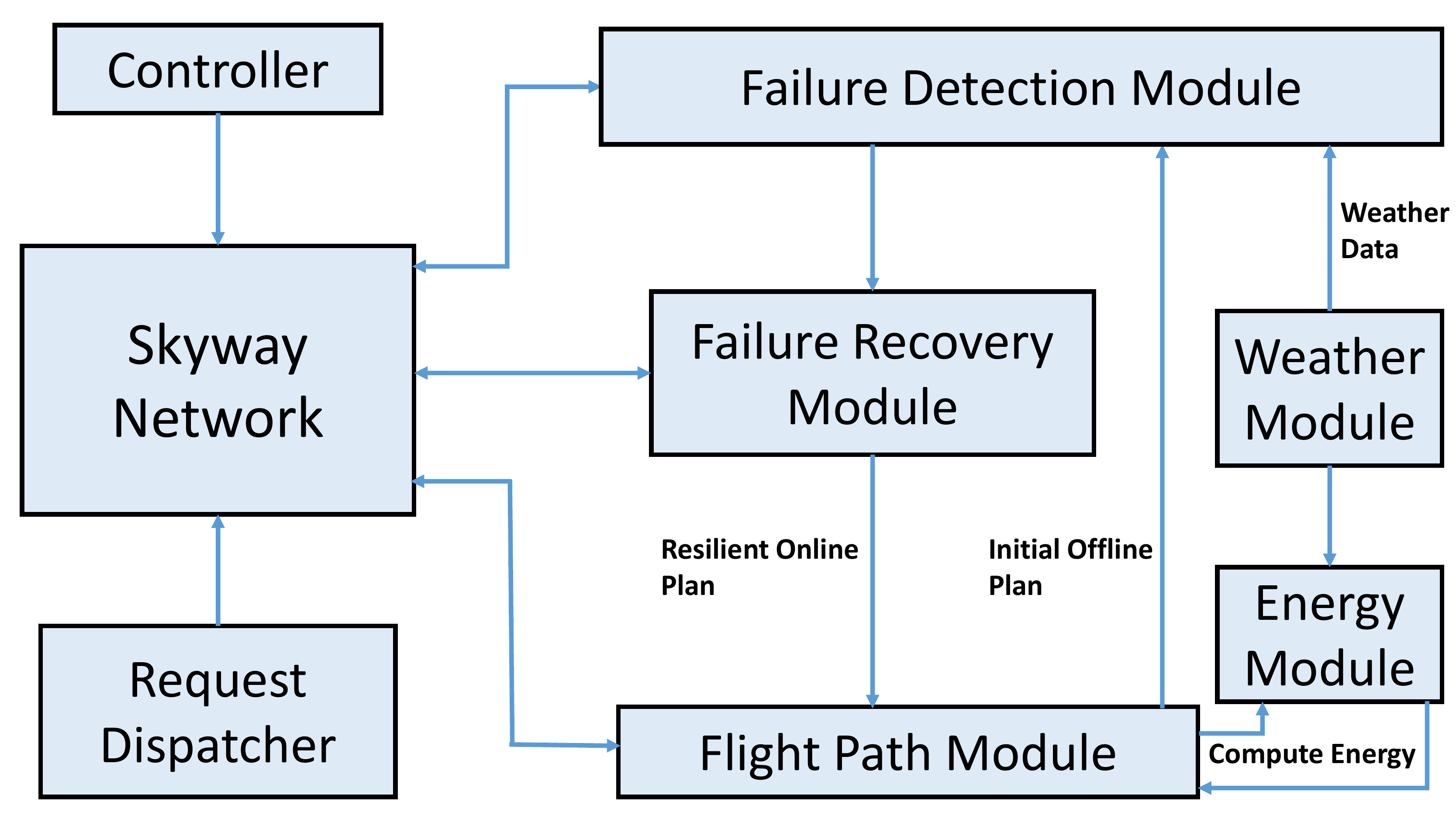}

    \caption{Structure of drone simulation model}

    \label{fig16}

\end{figure}

We design a custom drone simulation model using tools from drone energy consumption model \cite{9091085}, weather model \cite{roubeyrie2018windrose}, operations research, i.e., delivery service management \cite{hadjimichael2020rhodium}, and 2D path planning \cite{SciPyProceedings_11}. The simulation model consists of the following modules (as shown in Fig. \ref{fig16}): (a) \textit{controller}, (b) \textit{energy module}, (c) \textit{flight path module}, (d) \textit{weather module}, (e) \textit{request dispatcher}, (f) \textit{failure detection module}, (g) \textit{failure recovery module}, and (h) \textit{skyway network}. The controller module ensures the long-term stability of drone delivery services. It keeps track of all types of manoeuvres in the dynamic environment. The controller realizes the desired composition objectives by handling more and more services at each step. The energy module simulates the energy consumption of a drone service travelling from one recharging station to the next station. The energy consumption is calculated using the method in \cite{9091085}. An initial flight path is generated using our existing deterministic offline composition approach. The flight path module contains the composed services and position information of the drone services operating in the skyway network. The flight path is updated to maintain the resilience of composite services under dynamic weather conditions. The changing weather conditions influence the initial composition plan. The weather module is in charge of generating weather data for the whole simulation. The request dispatcher module takes care of receiving drone service requests from users. The current implementation of drone simulation model deals with single package delivery service request. The failure detection module monitors the execution of the initial composition plan. If a failure occurs due to dynamic weather conditions or stochastic arrival of other drone services, it notifies the failure recovery module. The failure recovery module is responsible for the execution of two main actions: (1) estimation of the failure impact and (2) local recomposition of affected drone services. As there is no 3D graphics involved (to simulate 3d drones), we do not require high capacity GPU. The simulation environment and composition algorithms are written in Python.

\begin{table}[t]
\centering
\caption{Dataset Description}

\label{tab:table2}
\begin{tabular}{|p{1.75cm}|p{6.75cm}|p{2.4cm}|}
\hline
 \textbf{Attribute} &  \textbf{Description} & \textbf{Example value}\\

\hline

Drone name &  Represents the manufacturer of the drone & DJI M200 V2 \tablefootnote{https://www.dji.com/au/matrice-200-series-v2/info\#specs} \\ \hline

Payload & Represents the weight a drone can carry (in kilograms) & 1.45 kg \\ \hline

Flight time & Represents the time a drone can fly with full payload capacity (in minutes) & 24 min \\ \hline

Range & Represents the distance a drone can cover with full payload capacity (in kilometres) & 32.4 km \\ \hline

Speed & Represents the flying speed of a drone with full payload capacity (in kilometres per hour) & 81 km/h \\ \hline

Recharging time & Represents the time required by a drone for recharging from 0\% to 100\% (in hours) & 2.24 hours \\ \hline


\end{tabular}
\end{table}

We use NetworkX \cite{SciPyProceedings_11} python library to construct the topology of the skyway network. We model the multiple delivery drones operating in the same skyway network. We evaluate the proposed approach using a real drone dataset \cite{14}. The dataset contains the trajectories of drones, which include data for coordinates, altitude, and timestamps. We augment a dataset for different types of drones considering the flight range, payload, battery capacity, speed, and recharging time. The details of the dataset are given in Table \ref{tab:table2}. The efficiency of the proposed framework depends on the values of the environmental variables. Table \ref{tab:table3} describes the environmental variables used in the experiments. The number of drones varies from 50-80 for varying sizes of the skyway network. We assume that each node is a recharging station. The number of nodes (i.e., recharging stations) varies from 10-60 for all approaches. Each recharging station has a finite number of recharging pads. The number of skyway segment DaaS services depends upon the size of the skyway network and the number of interconnected nodes. The proposed approach focuses on the single package delivery services from a given source to a destination. Each experiment starts with a random source and a destination point. The service failures occur randomly at runtime. The frequency of failures in each experiment varies from 10-50\% times the total number of nodes. The effect of each failure varies from a couple of subsequent nodes to the destination node. We conducted the experiments for 10\% times the total number of nodes and computed the average results.

\begin{table}[t]
\centering
\caption{Experiment Variables}
\label{tab:table3}
\begin{tabular}{|l|l|}
\hline
 \textbf{Variable} &  \textbf{Value} \\

\hline


Number of drones &  [50, 80]\\ \hline

Number of nodes (or recharging stations) & [10, 60] \\ \hline

Number of recharging pads at each station & 5 \\ \hline

Number of DaaS services & [500, 2500] \\ \hline

Number of generated requests & 1500 \\ \hline

Average battery consumption rate with 1 kg package & 25\%/10 km \\ \hline

Number of sources & 1 (random) \\ \hline

Number of destinations &  1 (random)  \\ \hline

Frequency of failures (\% times the total nodes) &  [10, 50] \\ \hline

Experiment run (\% times the total nodes) & 10 \\

\hline
\end{tabular}
\end{table}








\subsection{Baseline Approach}
To the best of our knowledge, this paper is the first attempt for a resilient drone service selection and composition in dynamic weather conditions. To evaluate our proposed approach, we compare the resilient drone service composition algorithm with Brute-Force algorithm. The Brute-Force approach is an all-paths search method. We apply the Brute-Force approach as a baseline to generate the ground truth of optimal compositions. We use Brute-Force in two phases of experiments to find optimal service compositions. In the first phase of experiments, Brute-Force approach finds all the possible compositions of drone services from a given source to a destination. We then select an optimal composition based on the QoS parameters of drone services. In the second phase of experiments, Brute-Force approach is used for the global recomposition of services to handle the service failures at runtime. Global recomposition refers to composing services from the failed point until the destination. Whenever a service failure occurs, Brute-Force approach finds all the possible compositions from current failed service until the destination. Finding all possible compositions of drone services is time exponential which is undesired. This significantly reduces the performance of Brute-Force approach to find optimal drone service composition and limits its use for large-scale problems.

\subsection{Without Lookahead Approach}

We use without lookahead approach in comparison to the proposed lookahead heuristic-based approach. The without lookahead approach behaves similar to a greedy shortest path algorithm. It always selects the least travel distance services leading towards the destination. The without lookahead approach has a higher probability to fail under dynamic weather conditions. For example, the initial composition plan and expected delivery time may be highly affected by adverse wind. Because of its greedy nature, the without lookahead is fast compared to baseline Brute-Force approach and the proposed lookahead approach. Sometimes, the selection of least travel distance services leads to the congested nodes which may result in longer delays for the availability of recharging pads.

\subsection{Results and Discussion}

The proposed approach performs composition of selective services based on certain parameters to reach the destination faster. We first generate an initial service composition plan and compare the Brute-Force, without lookahead, and lookahead approaches. We consider three evaluation parameters for comparison as follows: (1) average computation time, (2) average delivery time, and (3) average distance travelled. We then compare the Brute-Force and adaptive lookahead heuristic-based approaches dealing with the runtime service failures.

\subsubsection{Results for Initial Offline Service Composition}

\begin{figure}
    \centering
    \begin{minipage}{0.49\textwidth}
        \centering
        \includegraphics[width=\textwidth]{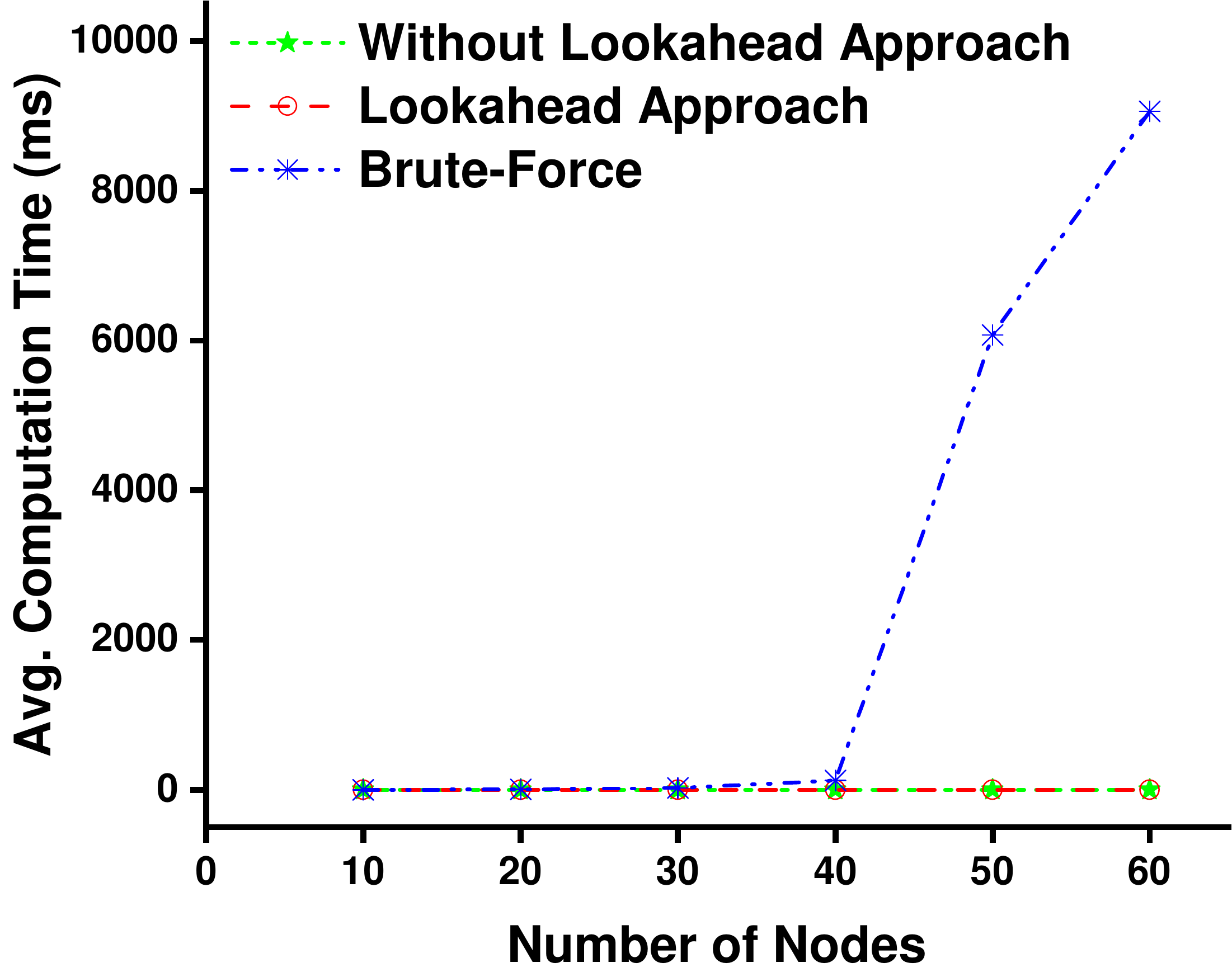} 
        \caption{Average computation time}
        \label{fig:fig7.pdf}
    \end{minipage}\hfill
    \begin{minipage}{0.49\textwidth}
        \centering
        \includegraphics[width=\textwidth]{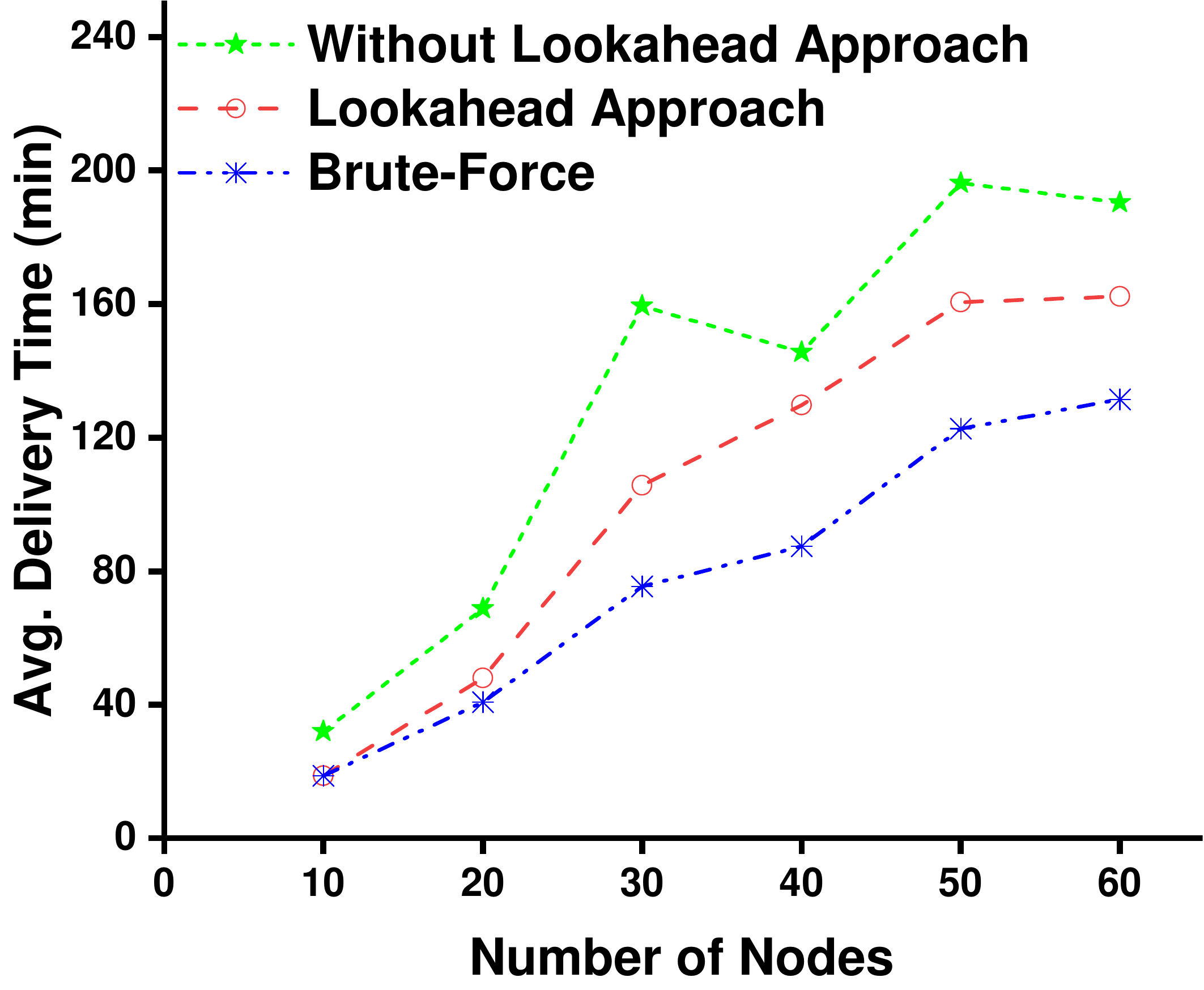} 
        \caption{Average delivery time}
        \label{fig:fig9.pdf}
    \end{minipage}
\end{figure}

\textbf{1) Average Computation Time:} The baseline Brute-Force approach is not time-efficient. The computational time for drone service composition using Brute-Force approach is very high in comparison to without lookahead and proposed lookahead heuristic-based approaches. The computation time increases due to the increasing number of possible compositions for drone services. Fig.~\ref{fig:fig7.pdf} compares the average computation time for Brute-Force, without lookahead, and proposed heuristic-based approaches. We observe that the proposed approach significantly outperforms the Brute-Force approach by drastically reducing the computational time. This is because the proposed approach avoids expensive computations by looking ahead once per neighbour state. The computation time varies for composing drone services depending upon the number of lookaheads. The higher the number of lookaheads we have, the more computational time is required to compose drone services.

\textbf{2) Average Delivery Time:} The delivery time of a drone service includes the flight time, recharging time, and waiting time. The selection of a \emph{right} drone service is of paramount importance as it ensures the availability of recharging pads ahead of time minimizing the overall delivery time. Fig.~\ref{fig:fig9.pdf} shows the efficiency of the proposed lookahead approach compared to Brute-Force and without lookahead approaches. The Brute-Force provides the exact solution as it finds all possible compositions. Our proposed approach obtains a near-optimal solution compared to Brute-Force approach. However, the time complexity of the proposed approach is much better than the baseline Brute-Force approach. Our proposed approach delivers the package 36\% faster than without lookahead approach. The without lookahead approach selects the services without anticipating the congestion conditions ahead which results in higher delivery time compared to our proposed approach. Our proposed approach uses a lookahead search strategy to reduce recharging and waiting times.

\begin{figure}
    \centering
    \begin{minipage}{0.49\textwidth}
        \centering
        \includegraphics[width=\textwidth]{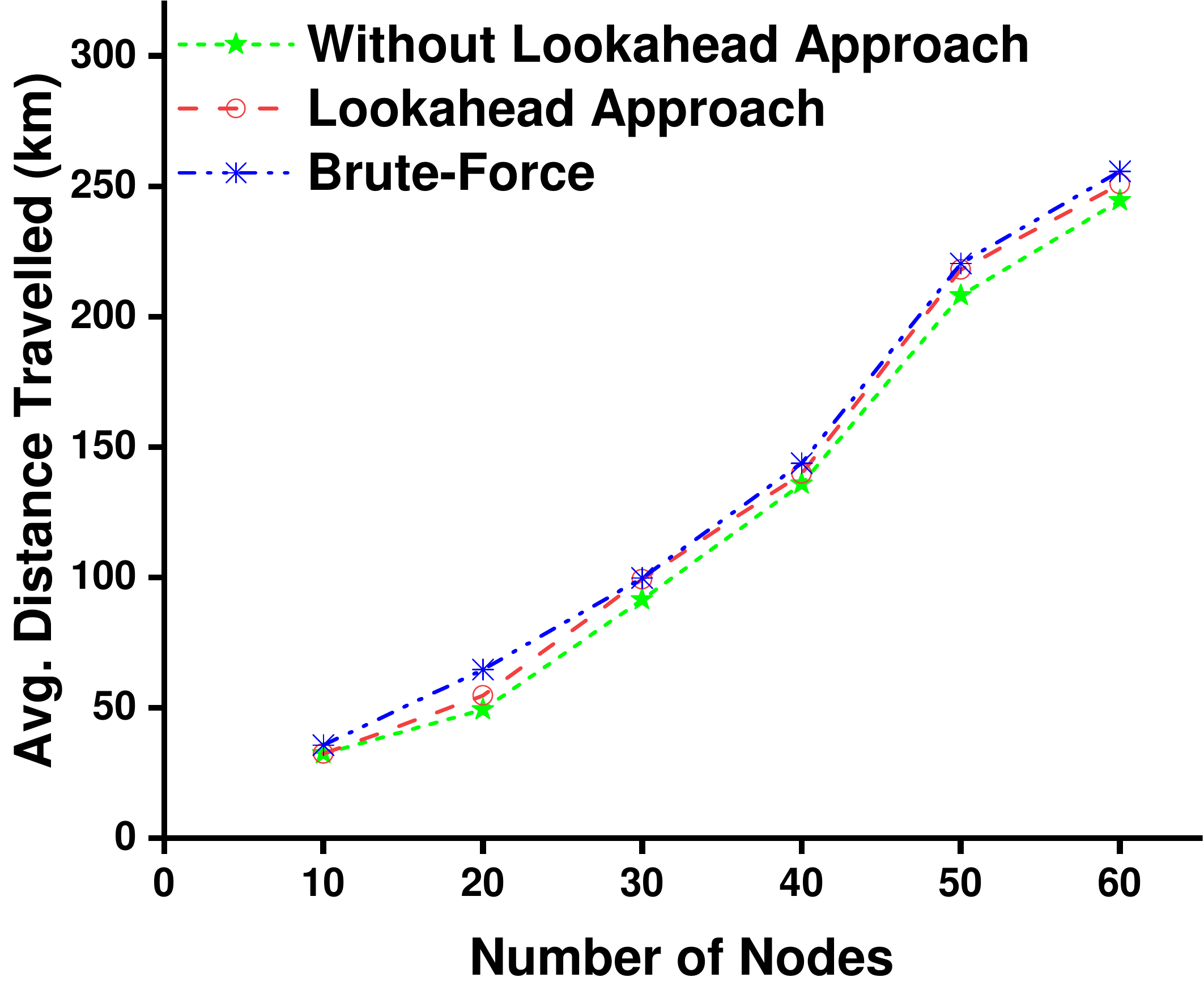} 
        \caption{Average distance travelled}
        \label{fig:fig11.pdf}
    \end{minipage}\hfill
    \begin{minipage}{0.49\textwidth}
        \centering
        \includegraphics[width=\textwidth]{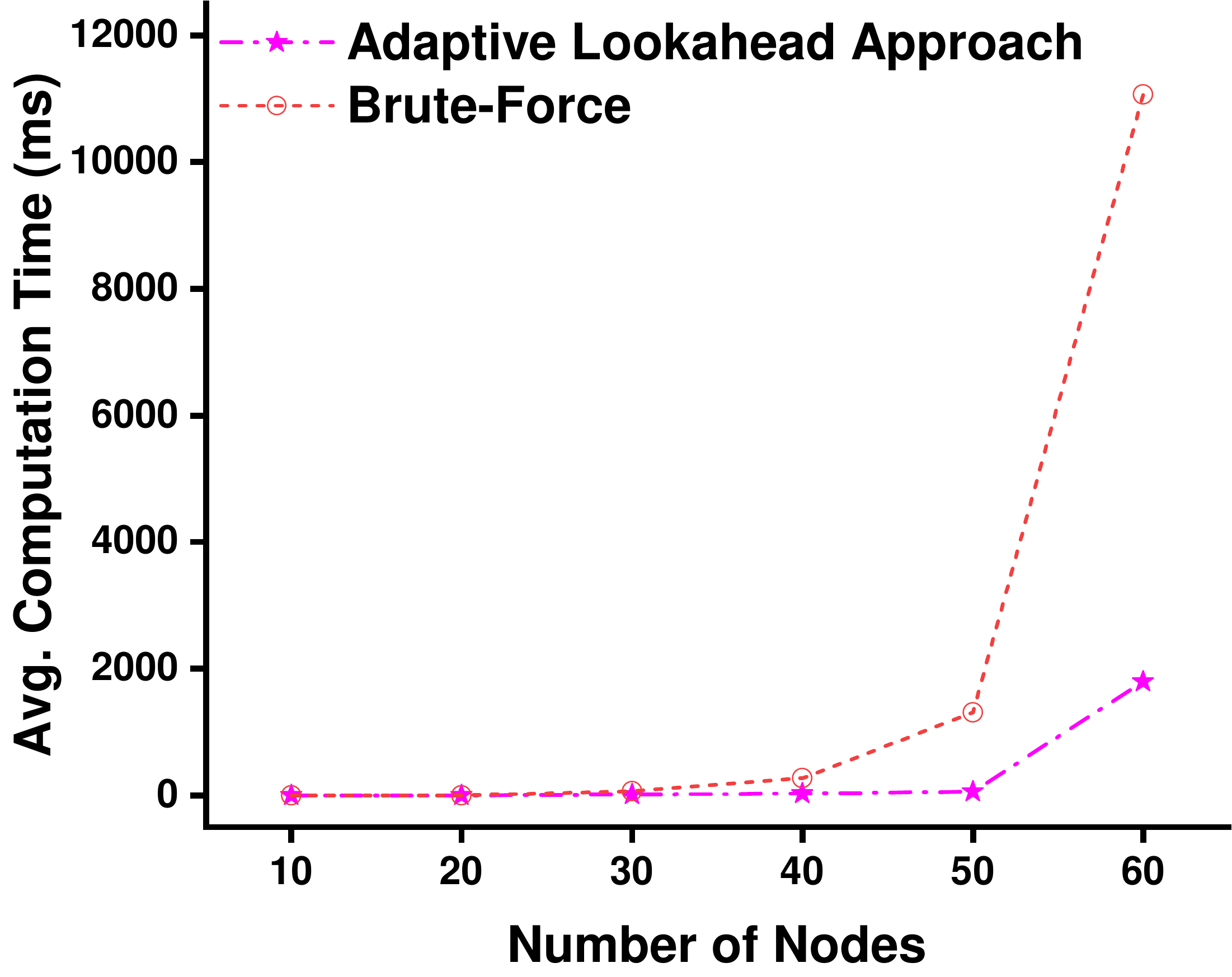} 
        \caption{Average computation time}
        \label{fig:fig8.pdf}
    \end{minipage}
\end{figure}

\textbf{3) Average Distance Travelled:} Some studies investigate the costs associated with drone delivery \cite{doi:10.1111/drev.10313}. The drone delivery cost for a package of 2 kg within a 10 km range is estimated at 10 cents in \cite{5}. For simplicity, we use the distance travelled by a drone as a cost function. Due to dynamic recharging constraints and wind conditions, the immediate drone services with least travel distance cost may lead to congested nodes. Fig.~\ref{fig:fig11.pdf} shows the average travel distances chosen by Brute-Force, without lookahead, and proposed heuristic-based approaches. The without lookahead approach always selects the least travel distance services, therefore, ends in higher delivery time. The Brute-Force approach always considers the least delivery time services leading towards the destination. Our proposed lookahead approach makes a decision based on next-to-adjacent node congestion information which results in 6\% improvement in delivery cost than the baseline approach.

\subsection{Results for Resilient Online Composition}

When a service failure occurs at any point during the execution, we recompose the services to meet the delivery demands. We use the Brute-Force approach for the global recomposition of drone service. While we propose adaptive lookahead heuristic-based local recomposition of affected drone services. In this context, global recomposition refers to the recomposition of services from failure point until the destination. The local recomposition refers to the recomposition of only the affected services in the initial composition plan.

\textbf{1) Average Computation Time:} The Brute-Force approach is highly time-consuming which is undesired. Whenever a failure occurs, it finds all possible compositions from failure point until the destination. The adaptive lookahead approach finds the best alternative composition from failure point until the next unaffected drone service in the initial composition or the next congested node. In the case of congested node selection, the failure effect on delivery time is compensated by subtracting it from waiting time at that node. Fig.~\ref{fig:fig8.pdf} plots the computation times of the baseline Brute-Force approach and the proposed heuristic-based approach. The computation time increases along with the number of services, which is an expected result. The computational complexity of our proposed approach is more consistent over time and less dependent on the network size. It is impractical to use the baseline approach in real-world scenarios as it is exhausted for large scale problems. 

\begin{figure}
    \centering
    \begin{minipage}{0.49\textwidth}
        \centering
        \includegraphics[width=\textwidth]{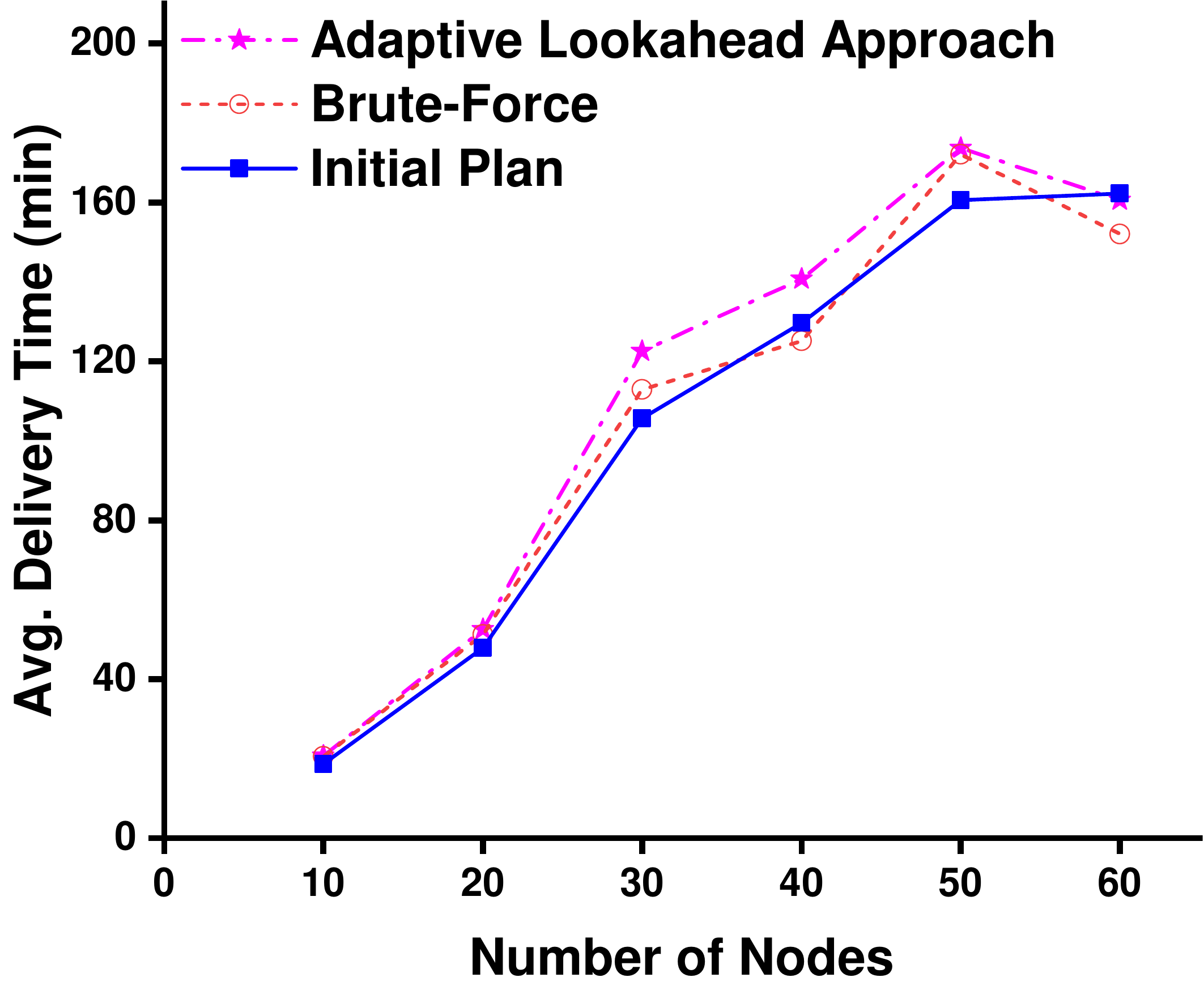} 
        \caption{Average delivery time}
        \label{fig:fig10.pdf}
    \end{minipage}\hfill
    \begin{minipage}{0.49\textwidth}
        \centering
        \includegraphics[width=\textwidth]{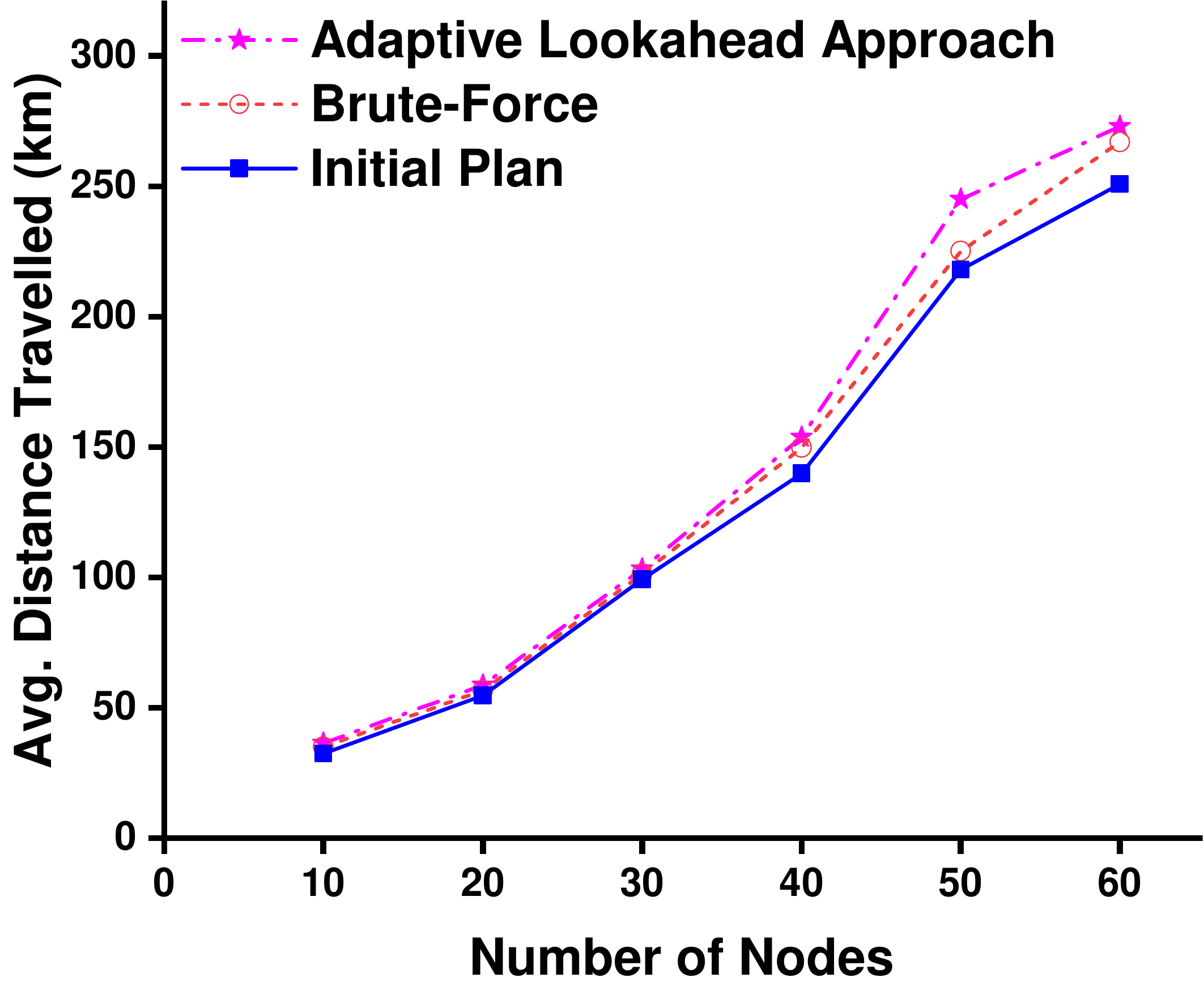} 
        \caption{Average distance travelled}
        \label{fig:fig12.pdf}
    \end{minipage}
\end{figure}

\textbf{2) Average Delivery Time:} The delivery time for drone services is highly uncertain when a single drone service cannot fulfil the user's requirements. The inter-dependencies on recharging constraints by other drones affect the overall delivery time of a drone service. At each station, the number of recharging pads are limited which can be occupied by other drones for long time periods. Fig.~\ref{fig:fig10.pdf} shows the comparison of the Brute-Force approach and the proposed approach compared to the initial plan. It shows that the local recomposition provides a near-optimal solution in a significantly shorter period of time compared to the Brute-Force approach. In some cases, only a single composition is possible from a failed point until the destination. In such cases, we simply replicate the delay effect of failures to the subsequent services. We observe that sometimes the Brute-Force approach finds better alternate composition than the original initial plan.

\textbf{3) Average Distance Travelled:} When a service failure occurs, the recomposition approach finds alternate routes to ensure the resilient delivery of drone services. In some cases, the travel distances may vary significantly compared to the original plan. Fig.~\ref{fig:fig12.pdf} plots the average travel distances chosen by Brute-Force and the proposed heuristic-based approaches on top of the initial composition plan. We observe that the performance of our proposed approach is almost linear up to 40 nodes in terms of travelling distance and maintains a notable trend even for a higher number of nodes. Our proposed approach saves a substantial amount of time to generate near-optimal solutions.

\begin{figure}
    \centering
    \begin{minipage}{0.49\textwidth}
        \centering
        \includegraphics[width=\textwidth]{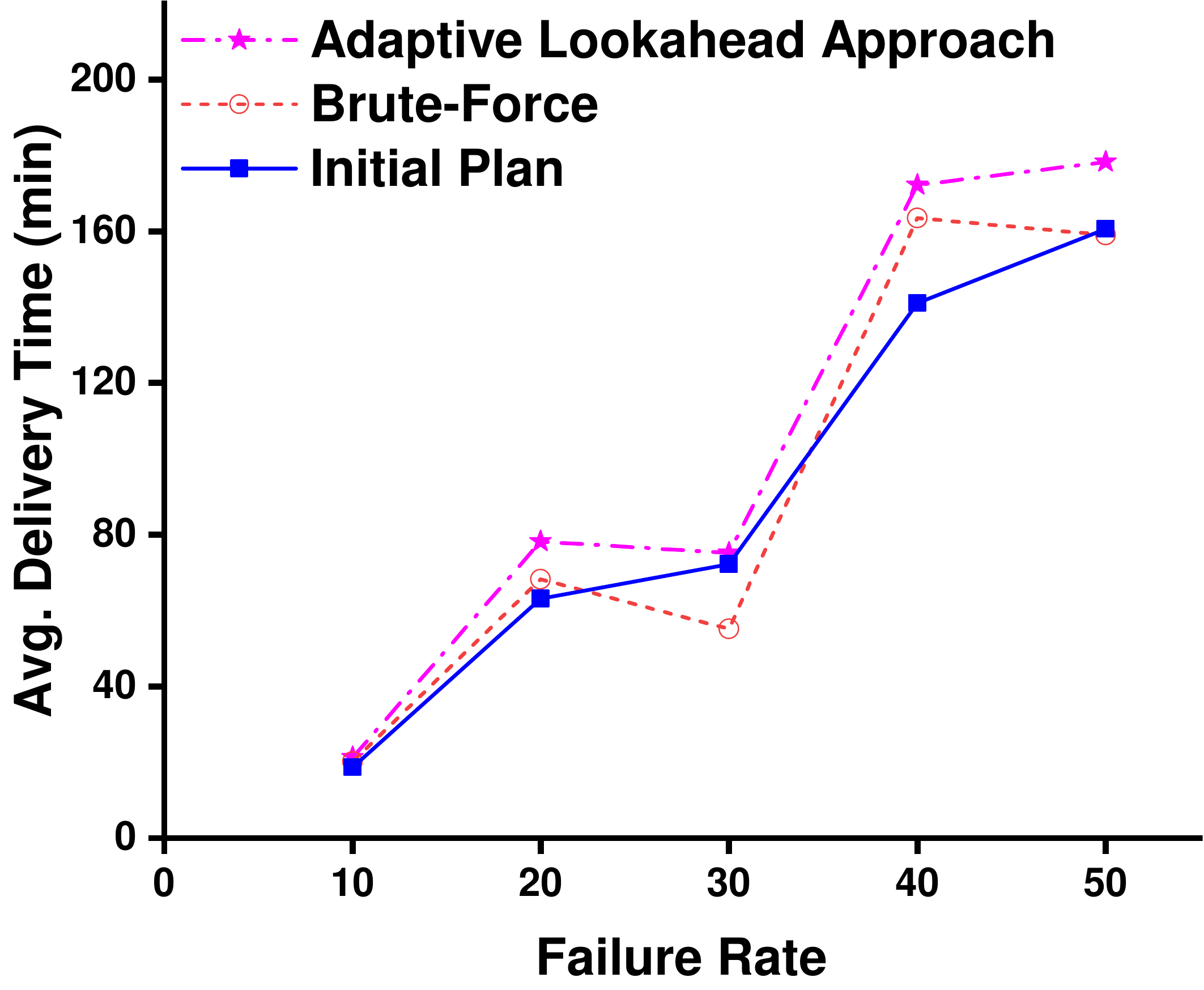} 
        \caption{Average delivery time}
        \label{fig:fig14.pdf}
    \end{minipage}\hfill
    \begin{minipage}{0.49\textwidth}
        \centering
        \includegraphics[width=\textwidth]{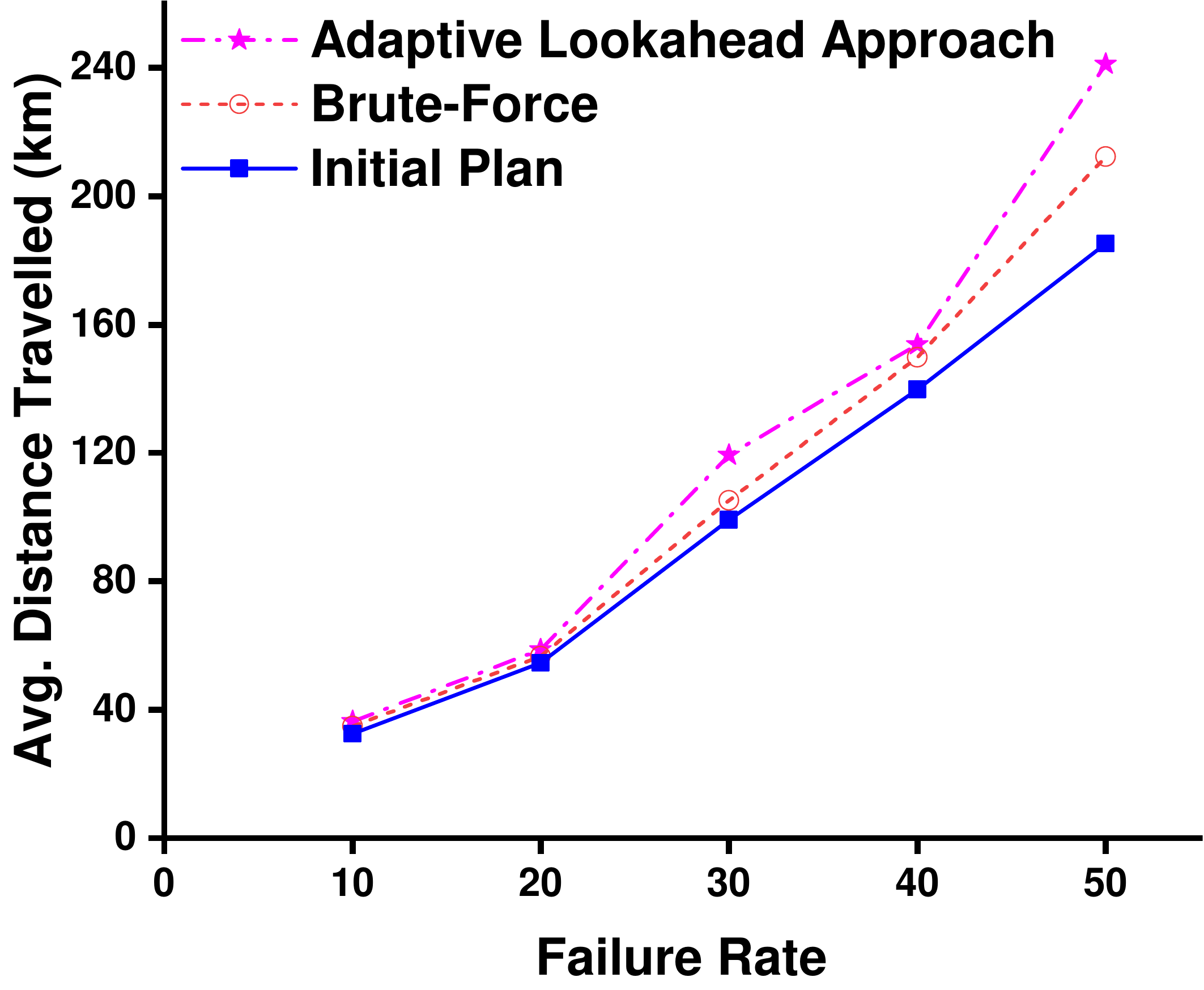} 
        \caption{Average distance travelled}
        \label{fig:fig15.pdf}
    \end{minipage}
\end{figure}

\textbf{4) Effects of Failure Rate:} We analyze the effects of the increasing number of failure rates on the resilience of delivery time and travel distance. Fig.~\ref{fig:fig14.pdf} and \ref{fig:fig15.pdf} plots the effects of different rates of failures on the average delivery time and distance travelled for the baseline Brute-Force approach and the proposed adaptive lookahead heuristic-based approach. We observe that the proposed approach finds optimal or near-optimal solutions for the increasing number of failure rates. The performance of our proposed approach is close to the Brute-Force approach even when the failure rate is high. Experiments based on the different failure rates demonstrate the effectiveness of our proposed approach in terms of delivery time and distance travelled (i.e., delivery cost).

\subsection{Lessons Learned}

We observed several unique features from our experiments with resilient composition during drone delivery operations. First, drones are vulnerable to weather conditions such as wind. The dynamic changes in the service environment may significantly influence the initial composition plan. Second, we may have a high failure rate of drone services due to dynamic weather conditions. The proposed adaptive recomposition algorithm can provide computationally efficient and near-optimal solutions in the dynamic environment. Moreover, the adaptive recomposition algorithm provides significantly better solutions when the number of services is small. Third, the computational complexity of the adaptive recomposition algorithm remains consistent even when the network size becomes large. Fourth, the use of local recomposition techniques over global recomposition techniques provides better practical solutions especially in terms of computational complexity. Finally, the use of global recomposition is impractical in real-world scenarios as it exhausts for large scale delivery networks.

\section{Conclusion}

We propose a resilient service composition framework for drone-based delivery considering the recharging constraints and dynamic weather conditions. An optimal set of candidate drone services is selected using the skyline approach at the source node in a skyway network. We present a formal model to represent constraint-aware drone services. We propose a deterministic lookahead algorithm to build an initial offline composition plan. We develop a heuristic-based resilient service composition algorithm that adapts to changes in service behaviour at runtime. We run several experiments to illustrate the performance of the proposed approach in comparison to Brute-Force and without lookahead approaches. We found that the proposed approach is runtime efficient and produces significantly better results than the Brute-Force and without lookahead approaches. Moreover, the proposed approach guarantees the resilience of delivery services for the increasing number of failure rates. Hence, it is a more practical solution in real-world applications of drone delivery services. 

A key limitation of the proposed approach is that the proposed approach does not take into account the handover of packages among different drones at intermediate recharging stations. The handover of packages to spare drones at intermediate recharging stations may assist in minimizing the overall delivery time. We plan to apply new optimization techniques for the handover initiation, the selection of the optimal drone service, and the handover management among different drones. The behaviour of a drone depends on wind patterns such as tailwinds and headwinds in different geographical areas. Another limitation of the proposed approach is that it does not incorporate the changing wind patterns into the drone service model. The proposed approach only focuses on the effects of wind speed and direction on the drone service composition plan. There are several weather conditions that can affect a drone's performance such as precipitation (e.g., rain, snow, hail, and sleet), temperature, cloud cover, and visibility. We intend to consider the effects of different weather conditions on the performance of a drone and exploit deep learning techniques for predicting and forecasting weather patterns. A single drone can deliver multiple small packages from a warehouse to desired destinations in one trip. The proposed resilient composition approach is limited to generate solutions for single package delivery by a drone from a given source to a destination. As future work, we plan to explore different adaptive techniques to extend the proposed approach for multi-package deliveries in a dynamic environment.

\subsubsection*{Acknowledgment}
This research was partly made possible by DP160103595 and LE180100158 grants from the Australian Research Council. The statements made herein are solely the responsibility of the authors.

\bibliography{mybibfile}

\end{document}